\documentclass[reprint,nofootinbib,amsmath,amssymb,fontenc aps]{revtex4-1}

\usepackage{graphicx,dcolumn,bm,hyperref,float}
\usepackage[mathlines]{lineno}

\newcommand{\sz}{0.6cm}

\begin{document}

\title{Statistics of Inflating Regions in Eternal Inflation}

\author{Mudit Jain$^1$}
\email{mudit.jain@tufts.edu}
\author{Mark P.~Hertzberg$^{1,2,3,4}$}
\email{mark.hertzberg@tufts.edu}
\affiliation{$^1$Institute of Cosmology, Dept.~of Physics and Astronomy, Tufts University, Medford, MA 02155, USA}
\affiliation{$^2$School of Physics, University of New South Wales, Sydney, NSW 2052, Australia}
\affiliation{$^3$School of Physics, University of Sydney, Sydney, NSW 2006, Australia}
\affiliation{$^4$Department of Physics, Tokyo Institute of Technology, Ookayama, Meguro-ku, Tokyo 152-8551, Japan}

\begin{abstract}
We compute the distribution of sizes of inflating and non-inflating regions in an eternally inflating Universe. As a first illustrative problem, we study a simple scenario of an eternally inflating Universe in the presence of a massless scalar field $\varphi$ whose field values lie within some finite domain $\varphi\in(-\varphi_{cr},\varphi_{cr})$, with $\pm\varphi_{cr}$ marking the onset of thermalization/crunching. We compute many important quantities, including the fractal dimension, distribution of field values among inflating regions, and the number of inflating and non-inflating Hubble regions. With the aid of simulations in 1 spatial dimension, we show this eternally inflating Universe reaches a steady state in which average sizes of inflating  regions grow only as a power law in the field's crunch value $\sim \varphi_{cr}^2$ (extension to higher dimensions is $\sim\varphi^{2/D}$), contrary to a naive expectation of an exponential dependence. Furthermore, the distribution in sizes exhibits an exponential fall off for large distances (with an initial power law for inflating regions). We leave other interesting cases of more realistic potentials and time varying Hubble parameter for future work, with a possible application to the SM Higgs in the early Universe.
\end{abstract}

\maketitle


\section{\label{sec:intro}Introduction}

Cosmological inflation is currently the best known explanation for large scale homogeneity and isotropy in our observable universe, as well as providing a detailed account of fluctuations \cite{Guth:1980zm,Linde:1981mu}. Eternally inflating scenarios, in which the Universe on the largest of scales continues to inflate forever, leading to a giant universe with both inflating and non-inflating patches, are quite generic and naturally occur in many inflationary models; see \cite{Linde:1983gd,Vilenkin:1983xq}.  It is possible then that our own observable Universe is one of these currently non-inflating (thermalized) regions within a giant eternally inflating Universe. A natural question to ask then is about the structure and distribution of inflating and non-inflating patches of such a giant Universe. For example it was shown in Ref.~\cite{Aryal:1987vn} that the eternally inflating Universe ultimately forms a fractal structure and  the fractal dimension was computed in various cases. Other important work includes Refs.~\cite{Linde:1993nz,Linde:1993xx}.

Important outstanding issues that remain are the sizes of inflating and thermalizing regions that would be present in this eternally inflating Universe. What are the typical sizes? What is their distribution? etc. We will address such questions in this work. An important scenario is that of an eternally inflating Universe in the presence of one or more scalar fields such that regions within some domain of field space continue to inflate, while regions in another domain of field space exit inflation. The field/s could either be the inflaton itself, or some other spectator field/s. For the former scenario, our observable Universe could be one of the thermalized regions surrounded by eternally inflating ones. The latter case could also be that of our own Universe with the scalar being the SM Higgs whose potential within the minimal SM can plausibly exhibit these different types of domains. That is regions in which the Higgs goes beyond the instability scale (arising due to quantum corrections from its interactions with other heavy particles like the top quark and the W, Z bosons \cite{Sher:1993mf,Casas:1994qy}) are marked as ``crunched" while regions in which it remains within the instability scale are still inflating\footnote{The question of Higgs instability and inflation in the early Universe has been studied in the literature; see \cite{Kearney:2015vba,East:2016anr,Kohri:2016wof,Espinosa:2015qea} and references therein.}. These regions can perhaps thermalize after a while either through some exit mechanism, or due to the inflaton decaying itself, in order for our observable Universe to be produced. For the present paper, we will not be focussed on the exact crunching/thermalizing mechanisms, but on the underlying structure of the eternally inflating Universe. Alterations and applications to more realistic models, and close connection to the SM Higgs, is left for future work.

In this paper, we analyze a scenario in which there is a single scalar field $\varphi$ that controls the eternal inflating dynamics. For simplicity, we take the field's potential to be constant $V(\varphi) = const$, with a constant Hubble $H$, within the field's domain $\varphi\in(-\varphi_{cr}, \varphi_{cr})$ (related work includes Ref.~\cite{Barenboim:2016mmw}). Here $\varphi_{cr}$ is some ``crunch" value, such that in the domain $|\varphi|>\varphi_{cr}$ inflation ends locally; here either the universe forms an AdS crunch, or perhaps just thermalizes and reheats, but in any case it no longer inflates. We refer to this as a ``cliff" potential. The global structure of the Universe is dictated by the motion of $\varphi$ as it determines whether local Hubble patches inflate or not. The specific model that approximates this dynamics is given ahead in eq.~\eqref{eq:langevin} for the regime in which the field is within $(-\varphi_{cr}, \varphi_{cr})$. Once it goes outside of this domain, the patch is declared to no longer inflate (for simplicity we will simply then take such a patch to become static, neither growing nor decaying, when we present the distribution of sizes of these non-inflating patches). So although any individual Hubble patch will eventually stop inflating, this basic process continues eternally throughout the Universe since daughter patches are always spawned at a sufficiently fast rate that some always survive and inflation never ends globally. This results in an assortment of ``regions" of collections of inflating Hubble patches surrounded by non-inflating ones at any moment in time. Furthermore, after waiting for some initial transient behavior to wash away, we are left in a steady state behavior of this variety. 

In this paper we determine the statistics for the sizes of inflating and non-inflating regions, along with other related quantities of interest like the fractal dimension, number density function for inflating patches etc, some of which have already been addressed in the literature \cite{Aryal:1987vn}. Also, we shall assume no drastic dynamics at the boundaries separating inflating and non-inflating regions. With the aid of simulations in 1-dimension, we shall show that the eternally inflating Universe gets to a steady state in which the total fractions of inflating and non-inflating regions become constants. The distributions in sizes exhibit exponential decay for large distances, while average sizes of inflating regions grow only as a power law ($\sim \varphi_{cr}^2$ and $\sim \varphi_{cr}^{2/D}$ in general). Furthermore, we develop a steady state, quasi-analytical, formalism to understand these results.. We leave more interesting cases of $V(\varphi) \neq const$ and in which classical drift can dominate over quantum fluctuations (semi-classical random walk), and a time dependent $H$ for future work, along with an application to the case of SM Higgs during inflation in the light of its instability at high energies.

The organization of the paper is as follows: In Section II, we review the well known stochastic approach towards a scalar field's dynamics in inflationary background. In Section III we point out that the eternally inflating Universe reaches a steady state, simultaneously calculating various quantities of interest. We also point out a kernel evolution technique, particularly useful to not only test the discrete simulations, but also provide a smooth interpolation to the continuous regime. In Section IV we describe two simulation networks that we used to model the scenario in 1-D, and compare the calculated quantities with them. In Section V and VI, we provide simulation results for the statistics of sizes of inflating regions and non-inflating regions respectively, along with their analytical understanding. In Section VII we summarize and discuss our findings. Finally, in the Appendix we provide some high order moments of the distribution.

\section{Stochastic Dynamics}

We shall adopt a sort of semi-classical approach in which one decomposes the scalar field into a low frequency part and a high frequency part and focus only on the low frequency part. In order to set up the problem, let us recap the dynamics of quantum fluctuations $\hat{\phi}(x)$ of a massless real scalar field ($V(\phi) = const$) in pure D-dimensional de-Sitter space ($D$ is the number of {\em spatial} dimensions) in which the Hubble parameter is $H = const$. Recall that the mode functions $\phi_{k}(t)$, which can be defined as usual via the following Fourier decomposition
\begin{equation}
\hat{\phi}(x,t) = \int \dfrac{d^Dk}{(2\pi)^D}\dfrac{e^{i{\bf{k}}.{\bf{x}}}}{\sqrt{2|{\bf{k}}|}}\,\phi_{k}(t)\,\hat{a}_{k} + \text{h.c}
\end{equation}
where $\hat{a}_{k}$ is the canonically normalized annihilation operator, obey the following equation of motion
\begin{equation}
\ddot{\phi}_{k} + D\,H\,\dot{\phi}_{k} + \dfrac{k^2}{a^2}\phi_{k} = 0.
\end{equation}
Here dots represents time derivatives and $a \propto e^{Ht}$ is the scale factor. This is a linear second order ODE and the solutions can be readily found for any arbitrary $D$ in terms of spherical Bessel functions. In 1 spatial dimension $D=1$, the solutions are simply sinusoids\footnote{We believe our results can be extended to higher dimensions in a straightforward fashion. Also our analytical estimates will be often performed with a general D. When we compare with 1 dimensional simulations we will set $D=1$.}. Requiring only positive frequency modes and choosing a unique vacuum (Bunch-Davies) as $t \rightarrow -\infty$ as usual, gives
\begin{equation}
\phi_{k}(t) = e^{-i\frac{k}{a H}}.
\end{equation}

Now, we may decompose the field $\hat{\phi}(x)$ into two pieces: a low frequency part, that we call $\hat{\varphi}(x)$, with all comoving modes in $(\varepsilon H,\varepsilon aH)$, and a high frequency part with all comoving modes greater than $\varepsilon aH$, and focus on the low frequency part only. Here $\varepsilon$ is an $\mathcal{O}(1)$ number that sets the cut-off scale. The two point correlation function is
\begin{eqnarray}
\langle\hat{\varphi}(x)\hat{\varphi}(y)\rangle &=& \int\limits_{{\varepsilon aH}>|k|>\varepsilon H}\dfrac{dk}{(2\pi)}\dfrac{|\varphi_{k}|^2}{2|k|}e^{i k\frac{(x-y)}{a}}\nonumber\\
&=& \dfrac{1}{2\pi}\left(\text{Ci}\left(r\varepsilon H\right) - \text{Ci}\left(\frac{r\varepsilon H}{a}\right)\right)
\label{eq:two.point.correlation}
\end{eqnarray}
and $|x-y| \equiv r$ is the \textit{physical} (not co-moving) separation. Here Ci is the cosine integral function: Ci$(x)\equiv-\int_x^\infty dt\cos(t)/t$. For super-horizon scales with $a\varepsilon^{-1}H^{-1} \gg r \gg \varepsilon^{-1}H^{-1}$ we have logarithmically dying out correlations
\begin{equation}
\langle\hat{\varphi}(x)\hat{\varphi}(y)\rangle = \dfrac{1}{2\pi}\log\left(\frac{a}{rH}\right) + O\left(1\right)
\end{equation}
while for sub horizon scales $r \ll \varepsilon^{-1} H^{-1}$ we have the random walk behavior
\begin{equation}
\langle\hat{\varphi}^2\rangle = \dfrac{H}{2\pi}t + O\left(r^2\right).
\end{equation}
One may also look at the mean field squared fluctuations $\langle\delta\hat{\varphi}^2\rangle (r) \equiv \langle\left(\hat{\varphi}(x)-\hat{\varphi}(y)\right)^2\rangle$, which according to the above can be written as
\begin{equation}
\langle\delta\hat{\varphi}^2\rangle (r) = \dfrac{2}{2\pi}\left(Ht - \text{Ci}\left(r\varepsilon H\right) + \text{Ci}\left(\frac{r\varepsilon H}{a}\right)\right),
\label{eq:mean.sq.theory}
\end{equation}
which goes to zero as $r \rightarrow 0$ (sub-horizon scales) and has a logarithmic growth $\sim \log(rH)$ for super-horizon scales (with $\mathcal{O}(1)$ correction factors). This behavior of correlations means that (the low frequency part of) field values in any two regions separated by some physical distance $r$ perform random walk together (when $r \lesssim H^{-1}$) until their separation is bigger than the horizon (when $r \gtrsim H^{-1}$), and after that they start to wander off in separate random walk trajectories\footnote{Notice that this behavior is qualitatively independent of the choice of cut-off $\varepsilon$, whose value will only be of relevance when we compare to discrete simulations.}. This well known stochastic random walk behavior of field fluctuations is quite useful in modeling semi-classical dynamics of a scalar field during inflation. In particular, under slow-roll approximation one can generalize the dynamics of these fluctuations for $V(\varphi) \neq const$ by adding a classical drift to the quantum diffusion, by using the Langevin equation:
\begin{equation}
\dfrac{d\varphi}{dN} + \dfrac{V'\left(\varphi\right)}{D H^2}= \kappa\,\eta_{N}
\label{eq:langevin}
\end{equation}
where $N$ is the number of e-foldings, $\eta_N$ is a random variable that exhibits Gaussian white noise ($\langle\eta_N\eta_{N'}\rangle = \delta\left(N-N'\right)$), and $\kappa$ is the random walk step size per e-folding (that we shall refer to as the ``kick").

\section{\label{sec:analytics}Steady state analysis of eternal inflation}

In addition to the Langevin equation \eqref{eq:langevin} describing field values in each Hubble patch, one can also write a differential (Fokker-Planck) equation for the \textit{number density} $\rho(\varphi,N)$ of inflating Hubble patches:
\begin{equation}
\dfrac{\partial\rho}{\partial N} = \dfrac{\partial}{\partial\varphi}\left[\dfrac{V'(\varphi)}{DH^2}\rho\right] + \dfrac{\kappa^2}{2}\dfrac{\partial^2\rho}{\partial\varphi^2} + D\rho
\label{eq:Fokker.Planck.modified}
\end{equation}
Note the additional term on the right hand side compared to the standard Fokker-Planck equation; this takes into account the changing normalization due to cosmic expansion. 

\subsection{Simplified Boundary Conditions:\\ Imposing Nodes}

Now this Fokker-Planck equation is only to apply within the field's domain $\varphi\in(-\varphi_{cr},\varphi_{cr})$, while near the boundaries, we anticipate a sudden change in behavior. This can perhaps be met by having nodes at $\pm\varphi_{cr}$ at all times, at least as a first approximation, as was advocated in Ref.~\cite{Vilenkin:1983xq}. We shall improve upon this approximation in the upcoming Section \ref{Kernel}. For our cliff potential $V'(\varphi)=0$ for $\varphi\in(-\varphi_{cr},\varphi_{cr})$, and for initial condition $\rho(\varphi,0) = \delta(\varphi)$, one obtains \cite{Aryal:1987vn}:
\begin{equation}
\rho(\varphi,N) = \dfrac{1}{\varphi_{cr}}\sum^{\infty}_{m=0}\cos\!\left(\dfrac{(2m+1)\pi\varphi}{2\varphi_{cr}}\right)
e^{\left(D-\frac{(2m+1)^2\pi^2\kappa^2}{8\varphi_{cr}^2}\right)N}
\end{equation}
giving the total number of inflating Hubble patches (integral from $-\varphi_{cr}$ to $\varphi_{cr}$) equal to 
\begin{eqnarray}
I(N) &=& \int^{\varphi_{cr}}_{-\varphi_{cr}}d\varphi\,\rho(\varphi,N)\nonumber\\
&=& \sum^{\infty}_{m=0}\dfrac{4}{(2m+1)\pi}e^{\left(D-\frac{(2m+1)^2\pi^2\kappa^2}{8\varphi_{cr}^2}\right)N},
\end{eqnarray}
which for large $N$ is dominated by the first term $m=0$.

Now let us define the inflating fractal dimension as
\begin{equation}
D_{F} \equiv {\ln I(N)\over N},
\end{equation}
which is the (normalized logarithm of) the total volume of the universe that is inflating in the model relative to the total volume that would be inflating if crunches were not allowed (so $D_F=D$ if crunching did not occur). At large $N$ this is
\begin{equation}
D_{F}(\infty) = D - \dfrac{\pi^2\kappa^2}{8\varphi_{cr}^2}
\label{eq:fractaldimension}
\end{equation}
and the normalized number density
\begin{equation}
\tilde{\rho}(\varphi,N) = \dfrac{\rho(\varphi,N)}{I(N)}
\end{equation}
goes to a constant function
\begin{equation}
\tilde{\rho}(\varphi,\infty) = \dfrac{\pi}{4\varphi_{cr}}\cos\left(\dfrac{\pi\varphi}{2\varphi_{cr}}\right).
\label{eq:steady.state.distribution}
\end{equation}
Note that we evidently need to have $\varphi_{cr} \ge \pi\kappa/(2\sqrt{2D})$ in order for eternal inflation to occur with $D_F>0$. For the critical value $\varphi_{cr}=\pi\kappa/(2\sqrt{2D})$, we have an equilibrium situation where there is no net inflation and the amount of inflating regions doesn't change with time.

Now since we know the total number of inflating Hubble patches as a function of $N$, we can also calculate the number that have exited: For a small discretization step $\epsilon$, the number of regions $\bar{I}$ that exited inflation in between $N$ and $N+\epsilon$, is equal to the difference of the number of regions that would have inflated in between $N$ and $N+\epsilon$ if there were no exiting, with that of the actual number that did inflate in between $N$ and $N+\epsilon$. That is, we have the following simple differential equation
\begin{equation}
\dfrac{d \bar{I}}{d N} = DI - \dfrac{d I}{d N}.
\end{equation}
Note the implicit assumption made here that exited Hubble patches just accumulate as time goes on, without expanding. With the initial condition $\bar{I}(0) = 0$, we have
\begin{eqnarray}
\bar{I}(N) =&& \sum^{\infty}_{m=0}\dfrac{4(2m+1)\pi\kappa^2}{\left(8D\varphi_{cr}^2-(2m+1)^2\pi^2\kappa^2\right)}\times\nonumber\\
&&\;\;\;\;\;\;\;\;\;\;\;\;\left[e^{\left(D-\frac{(2m+1)^2\pi^2\kappa^2}{8\varphi_{cr}^2}\right)N}-1\right].
\end{eqnarray}
With both $I(N)$ and $\bar{I}(N)$ increasing with the same exponentials, the fraction of inflating and non-inflating patches
\begin{eqnarray}
f = \dfrac{I(N)}{I(N)+\bar{I}(N)}, \;\;\;\;\;\;\;\bar{f} = \dfrac{\bar{I}(N)}{I(N)+\bar{I}(N)},
\end{eqnarray}
become constants eventually, and we have
\begin{eqnarray}
f(\infty) = 1 - \dfrac{\pi^2\kappa^2}{8\varphi_{cr}^2D}, \;\;\;\;\;\;\;\bar{f}(\infty) = \dfrac{\pi^2\kappa^2}{8\varphi_{cr}^2D}.
\label{eq:fractions}
\end{eqnarray}
Since they approach constants at late times, we establish steady state behavior. The meaning of this can perhaps be best visualized in comoving coordinates (with scale factor $a\propto \exp(D_{f}(\infty) N/D)$) where this eternally inflating Universe is in equilibrium between creating regions that inflating and creating regions that exit. Note that the fraction of inflating patches does not go to unity exponentially fast with increasing $\varphi_{cr}$, which will be important for understanding the statistics for the sizes of inflating regions. Furthermore, it's important to point out that the number of e-foldings it takes to convergence towards steady state is quadratic in $\varphi_{cr}/\kappa$. Another important feature is that this steady state behavior does not depend sensitively on the choice of initial conditions; as long as the initial field value is far enough from $\varphi_{cr}$ so that eternal inflation is possible, we are guaranteed to approach this steady state.

\subsection{\label{Kernel}Accurate Boundary Conditions:\\ Kernel Propagation} 

Now to improve upon our assumption of having nodes at $\varphi=\pm\varphi_{cr}$, we point out a simple yet very useful technique to get these fractions and normalized number density function, which will also provide an interpolation between the discrete simulation and the continuous theory.

Firstly, let us construct a version of the Langevin equation (eq.~\eqref{eq:langevin}) that is discretized in time, with time step $\delta t=\epsilon$, as
\begin{equation}
\dfrac{\varphi_{n} - \varphi_{n-1}}{\epsilon} + \dfrac{V'\left(\varphi_{n-1}\right)}{D H^2} = \kappa\,\eta_{n},
\label{eq:langevindiscrete}
\end{equation}
with $\eta_{n}$ a random variable implementing a discrete version of quantum diffusion. It has the following Gaussian probability density
\begin{equation}
P\left(\eta_{n}\right) = \sqrt{\dfrac{\epsilon}{2\pi}}\,e^{-\frac{\epsilon}{2}\eta_{n}^2}.
\end{equation}
In general we can construct the kernel that propagates the (discretized) number density $\rho_{n}$:
\begin{equation}
K\left(\varphi_{n},\varphi_{n-1},\epsilon\right) = \dfrac{e^{D\epsilon}}{\sqrt{2\pi\kappa^2\epsilon}}\,e^{-\frac{\epsilon\left(\frac{\varphi_{n}-\varphi_{n-1}}{\epsilon} + \frac{V'\left(\varphi_{n-1}\right)}{DH^2}\right)^2}{2\kappa^2}},
\label{eq:kernel}
\end{equation}
with no support outside of the domain $(-\varphi_{cr},\varphi_{cr})$. This leads to the recursion relation
\begin{eqnarray}
\rho_{n}\left(\varphi_{n},n\epsilon\right) &=& \int^{\varphi_{cr}}_{-\varphi_{cr}} d\varphi_{n-1}K\left(\varphi_{n},\varphi_{n-1},\epsilon\right)\rho_{n-1}\left(\varphi_{n-1}\right)\nonumber\\
&=& \prod^{n-1}_{i=0}\int^{\varphi_{cr}}_{-\varphi_{cr}}d\varphi_{i}K\left(\varphi_{i+1},\varphi_{i},\epsilon\right)\rho_{0}\left(\varphi_0,0\right).\,\,\,\,\,\,\,\,\,\,\,
\end{eqnarray}
This is an iterative matrix multiplication technique (upon discretization of field space) and will converge to the dominant eigenstate of the kernel (the steady state number density function). Also, the total number of inflating and non-inflating patches at any step $n$ of the iteration is
\begin{eqnarray}
I_{n} &=& \int^{\varphi_{cr}}_{-\varphi_{cr}} d\varphi_{n}\rho_{n}(\varphi_{n}),\\
\bar{I}_{n} &=& \bar{I}_{n-1} + e^{D\epsilon}I_{n-1} - I_{n},
\label{eq:inflating.crunched}
\end{eqnarray}
with initial values $I_{0} = 1$, and $\bar{I}_{0} = 0$. From this we can easily get the fractions $f_{n}, \bar{f}_{n}$ and the fractal dimension $D_{F_{n}}$ at any step: 
\begin{eqnarray}
f_{n} &=& \dfrac{I_{n}}{I_{n}+\bar{I}_{n}},\\
\bar{f}_{n} &=& \dfrac{\bar{I}_{n}}{I_{n}+\bar{I}_{n}},\\
D_{F_{n}} &=& \dfrac{1}{n\epsilon}\ln\left(I_{n}\right).
\label{eq:goodnbadnumberdiscrete}
\end{eqnarray}
For the simple model here, $V = const$ giving a relatively simple random walk kernel.

\section{\label{sec:simulationsetup}Simulated Networks}

We now describe our simulation setup in order to numerically see the evolution. We work with two different networks to simulate inflation in 1 spatial dimension. The second network will be the same as that of Aryal and Vilenkin \cite{Aryal:1987vn}. The kick is $\kappa = 1/\sqrt{2\pi}$ in 1-D and in both networks we double the size of physical Universe at every step (a 2-folding with $\epsilon  =\ln 2$). The Langevin equation for each Hubble patch (with $V = const$) is then
\begin{equation}
\varphi_{i} = \varphi_{i-1} + \sqrt{\dfrac{\ln 2}{2\pi}}\,\zeta_{i} \equiv \varphi_{i-1} + \delta\varphi_{i}
\label{eq:Langevin.simulation}
\end{equation}
with $\zeta_{i}$ a Gaussian random variable with variance $=1$. We generate random variables $\delta\varphi$'s at each step in order to update the field values in each ``cell" (representative of a Hubble patch). In ``network-I", we have a literal expansion of the Universe in which we create more and more cells at every step, while ``network-II" essentially involves unfolding a comoving volume that is to become the final Universe at the end. The precise construction and details of these networks is explained next.

\subsection{Network-I}

\begin{figure}[t]
\centering
\includegraphics[width=1\columnwidth]{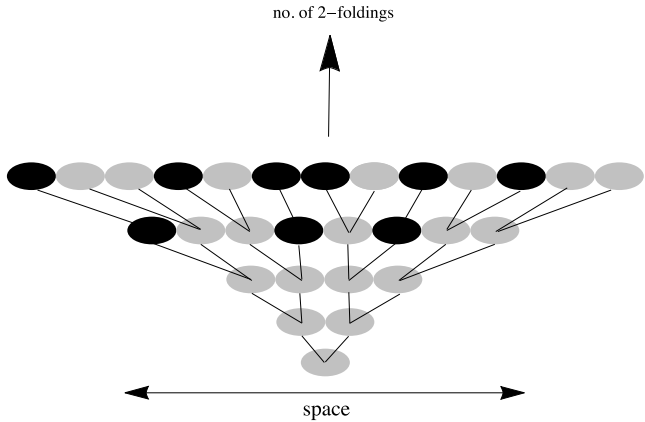}
\caption{A pictorial representation of network-I for 4 2-foldings. Grey cells correspond to inflating Hubble patches, while black cells correspond to ones that have exited.}
\label{Network1Pictorial}
\end{figure}

In this setup, we start with a single cell with field value $\varphi=0$, and then double the number of cells at the next step (2-folding) with the field values in the two daughter cells set by two (independent) $\delta\varphi$'s according to the above Langevin equation. Then at the next step each of the two cells spawn two new daughter cells with their field values chosen according to the same prescription, and so on. If at any step a cell goes outside the domain $(-\varphi_{cr},\varphi_{cr})$, it crunches/thermalizes, then we just keep it as it is in the next step. A pictorial representation of network-I is given in Fig.~\ref{Network1Pictorial}.

\begin{figure}[t]
\centering
\includegraphics[width=1\columnwidth]{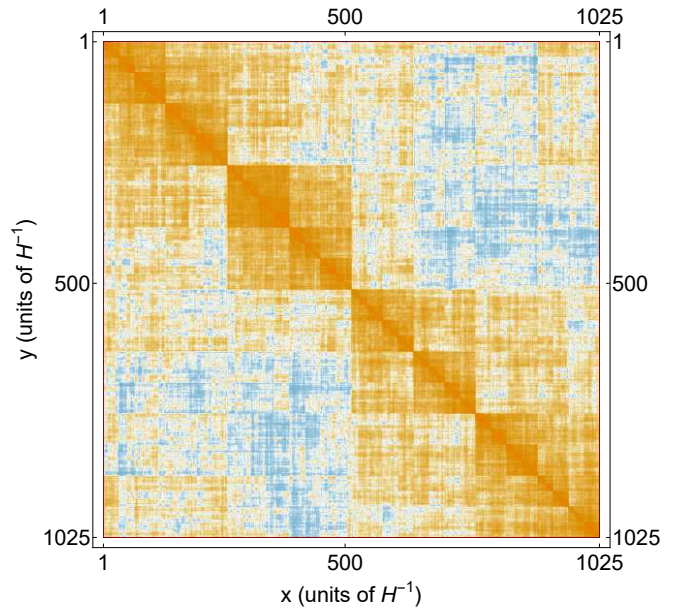}
\caption{A color coded plot of the ensemble averaged (over 100 independent runs) two point correlation function $\langle\varphi(x)\varphi(y)\rangle$ for network-I with $10$ 2-foldings and no thermalization/crunching. It is evident that the correlation function breaks statistical translation invariance as it has a dependence on $x + y$ (upper right to lower left diagonals), and the branching tree-like structure is apparent from the presence of square boxes.}
\label{Network1correlation}
\end{figure}

Now a complete model of inflation would exhibit statistical translation invariance, since the Bunch-Davies wave-function does so. So although any particular realization of the universe breaks translation invariance, it is recovered in an ensemble averaged sense. However, this particular network model doesn't maintain translational invariance even after ensemble averaging due to its tree like structure. This can be seen clearly from the ensemble averaged two point correlation function $\langle \varphi(x)\varphi(y)\rangle$ shown in Fig.~\ref{Network1correlation} (ignoring crunching for simplicity).

However, for a given observer, they obviously do not see the ensemble average anyhow. Instead at best one can compare volume averaging to observations. Here we will show that for volume averaging, correlations are in fact in rather good agreement with theory. At any step $n$ in the simulation, for a given distance $r$ (an integer in units $H=1$) there exists a $z_{0} = \text{Ceil}\left(\log_{2}(1 + r)\right)$ such that for every integer $z$ in $[z_{0},z_{0}+1,z_{0}+2,..,n]$, the number of pairs of cells ($r$ apart) with mean field squared fluctuations equal to $2\kappa^2\epsilon z$ is $\{(2^{z_0}-r)2^{n-z_{0}}, r\,2^{n-z_{0}-1}, r\,2^{n-z_{0}-2},..,r\}$ respectively. Therefore, on average we have the following mean field squared fluctuations
\begin{eqnarray}
\langle\delta\varphi^2\rangle (r) &=& \dfrac{2\kappa^2\epsilon}{2^{n}-r}\left[(2^{z_0}-r)2^{n-z_{0}}z_{0} + r\sum^{n}_{z=z_{0}+1}2^{n-z}\right]\nonumber\\
&=& \dfrac{2\kappa^2\epsilon}{2^{n}-r}\left[2^{n}z_{0} + r\left(2^{n-z_{0}+1}-n-2\right)\right]
\end{eqnarray}
which for large $n$ limit gives
\begin{equation}
\langle\delta\varphi^2\rangle (r) =\dfrac{2}{2\pi}\epsilon\,z_{0} + O(1).
\label{eq:two.point.vol.avg}
\end{equation}
For $r=0$ (being equivalent to the sub-horizon scale), we have zero since all cells undergo random walk, while for $2^{n} \gg r \gg 1$ (being equivalent to super-horizon scale) we have a log correlation. This is in qualitative agreement with the continuum behavior of the full theory (eq.~\eqref{eq:mean.sq.theory}) and proves that network-I has the right structure for volume averaging. Also, in order to match the network's correlations \textit{exactly} with the theory, we must make the right choice for the cut-off $\varepsilon$ which we can simply obtain by matching the $\mathcal{O}(1)$ correction terms; we get $\varepsilon = e^{2\epsilon - \gamma} = 4/e^{\gamma}$ where $\gamma$ is the Euler's constant. Ahead in Fig.~\ref{VolumeAverageCorrelation} we show the volume averaged two point correlation function along with the theory result with this chosen $\varepsilon$, showing good agreement. This exact matching however, is not qualitatively important for our purposes since it can be roughly captured by an $O(1)$ scaling of $r$ without changing the overall functional dependence of volume averaged quantities that are of interest. 

On the other hand, one may be concerned that volume averaging was required to enforce translational variance in this network. For this reason we also work with another simulation setup which improves upon this, which we describe next.

\subsection{Network-II}

\begin{figure}[t]
\centering
\includegraphics[width=1\columnwidth]{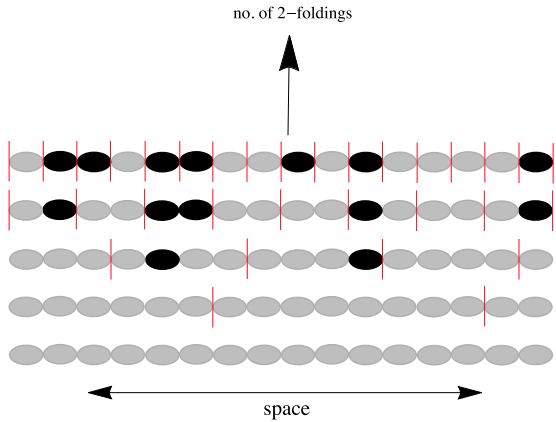}
\caption{A pictorial representation of network-II for $4$ 2-foldings. Grey cells correspond to inflating comoving patches, while black cells correspond to ones that have exited. Red lines mark the boundaries of h-regions.}
\label{Network2Pictorial}
\end{figure}

Our second simulation setup is the same as that of Ref.~\cite{Aryal:1987vn} in 1 spatial dimension. We begin with an array of $2^{n}$ cells where $n$ is the total number of 2-foldings that the simulation will run for, and with field values equal to zero in all of them. Then at the first step we choose an arbitrary lattice site as the origin and divide the array into ``h-regions" (in the language of Ref.~\cite{Aryal:1987vn}) of size $2^{n-1}$ each. We generate independent $\delta\varphi$'s for each h-region with every cell in it updated by the same $\delta\varphi$, and so on. At any $k^{th}$ step we create h-regions of sizes $2^{n-k}$ from an arbitrary chosen lattice site and generate those many $\delta\varphi$'s, with cells in a h-region updated with the same $\delta\varphi$. h-regions at the boundaries are generally not complete. At the last step we have $2^n$ h-regions i.e. each cell is an h-region and is updated by its own $\delta\varphi$. In any step if a cell thermalizes, we don't update it's field value anymore and just keep the value as it is in further steps. A pictorial representation of network-II is given in Fig.~\ref{Network2Pictorial}.

\begin{figure}[t]
\centering
\includegraphics[width=1\columnwidth]{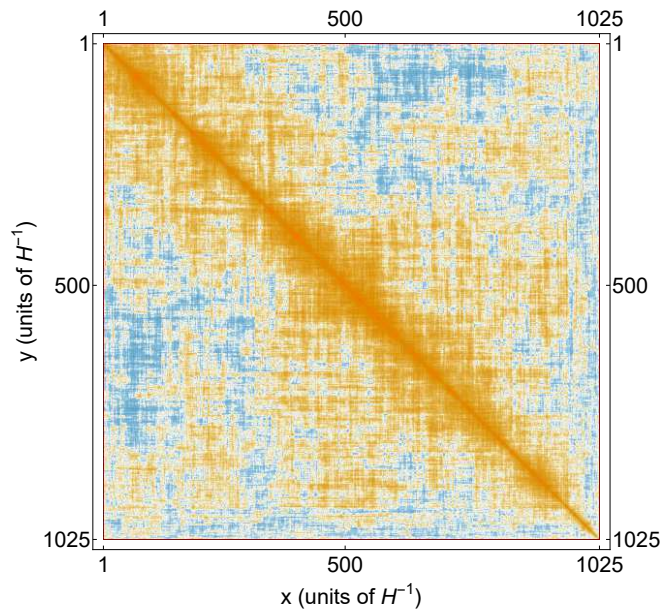}
\caption{A color coded plot of ensemble averaged (100 independent runs) two point correlation function $\langle\varphi(x)\varphi(y)\rangle$ for network-II with $10$ 2-foldings and no thermalization/crunching. This evidently improves upon translation invariance relative to the network-I result of Fig.~\ref{Network1correlation}.}
\label{Network2correlation}
\end{figure}

This network enjoys statistical translational variance (since we don't have a fixed center) which can be seen from the ensemble averaged two point correlation function shown in Fig.~\ref{Network2correlation}. However, for volume averaging it will always be less efficient than network-I.
This is because we need to choose a random origin at each step, and for any given distance $r$ this network can never produce more correctly correlated pairs ($(2^{z_0}-r)2^{n-z_{0}}$) than the previous network. It can only match it for a run where at every step there happens to be a boundary of a h-region at the center. For a typical run, this will not be the case. This difference, however, will only be prominent for smaller $r$'s. While for larger $r's$ it will have the same qualitative structure. Therefore this network will also serve our purposes for volume averaged quantities. 

\begin{figure}[t]
\centering
\includegraphics[width=1\columnwidth]{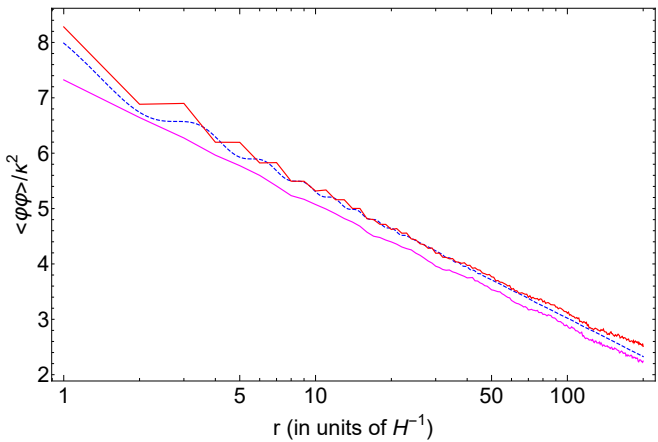}
\caption{Plot of volume averaged two point correlation functions $\langle\varphi\varphi\rangle(r)$ after $13$ $2$-foldings (with no thermalization/crunching) and averaged over $100$ independent runs. Dashed blue is the theory result eq.~\eqref{eq:two.point.correlation} with $\varepsilon = 4/e^{\gamma}$, while red and magenta is for network I and II respectively.}
\label{VolumeAverageCorrelation}
\end{figure}

Fig.~\ref{VolumeAverageCorrelation} also shows volume averaged two point correlation function of network-II (along with network-I). Note that due to this added randomness of the choice of origin, in order for this network's correlations to \textit{exactly} match the analytical result, we need to have a different $\varepsilon$ than before. However, as mentioned earlier, this is not qualitatively important and we put these details aside. Also, since we start with the total amount of cells and zoom in at every step in powers of $2$ irrespective of the arrangement of thermalized/crunched cells, we formally inflate these regions as well. However, this is only by way of appearance and has no consequence for the statistics of the non-thermalized/non-crunched regions.

Our final results are very similar for both networks upon volume averaging. The objection towards the first not maintaining translational invariance exactly is not so important since we are always interested in volume averaged quantities from the point of view of a single observer anyhow.

\subsection{Comparison with simulations}

Having discussed both networks, we now show various plots of the above calculated quantities for $D=1$ and compare with simulations. For each of the two networks, we have simulated 20 independent Universes for $\varphi_{cr}$'s (in units of $\kappa = 1/\sqrt{2\pi}$) in between $1$ and $10$, and up to $18$ 2-foldings. From kernel propagation, we obtain two sets of curves, one to compare with simulations (usually also evolved up-to the corresponding number of 2-foldings as simulations) and another to interpolate between $\epsilon = \ln 2$ and $\epsilon = 0$ at steady state. To obtain steady state behavior from kernel propagation, we compare the norm of the normalized number density $\tilde{\rho}$ in between two successive steps until they are different by 1 part in $10^{3}$. Also, we have adopted a convention of keeping green color for curves obtained from kernel propagation while blue for the analytical results. Red and magenta are for simulation networks-I and II respectively.

\begin{figure}[t]
\centering
\includegraphics[width=1\columnwidth]{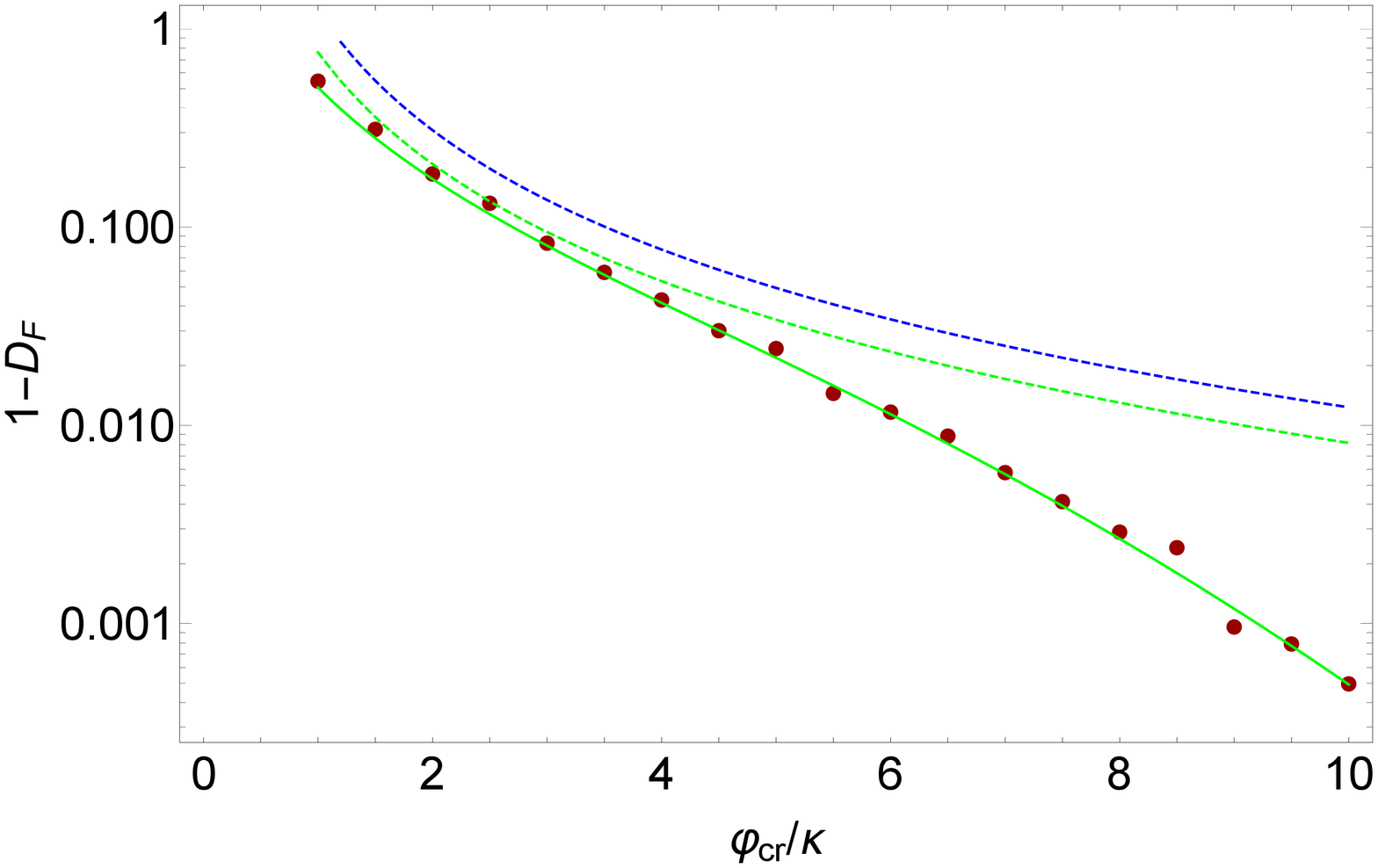}\vspace{\sz}
\includegraphics[width=1\columnwidth]{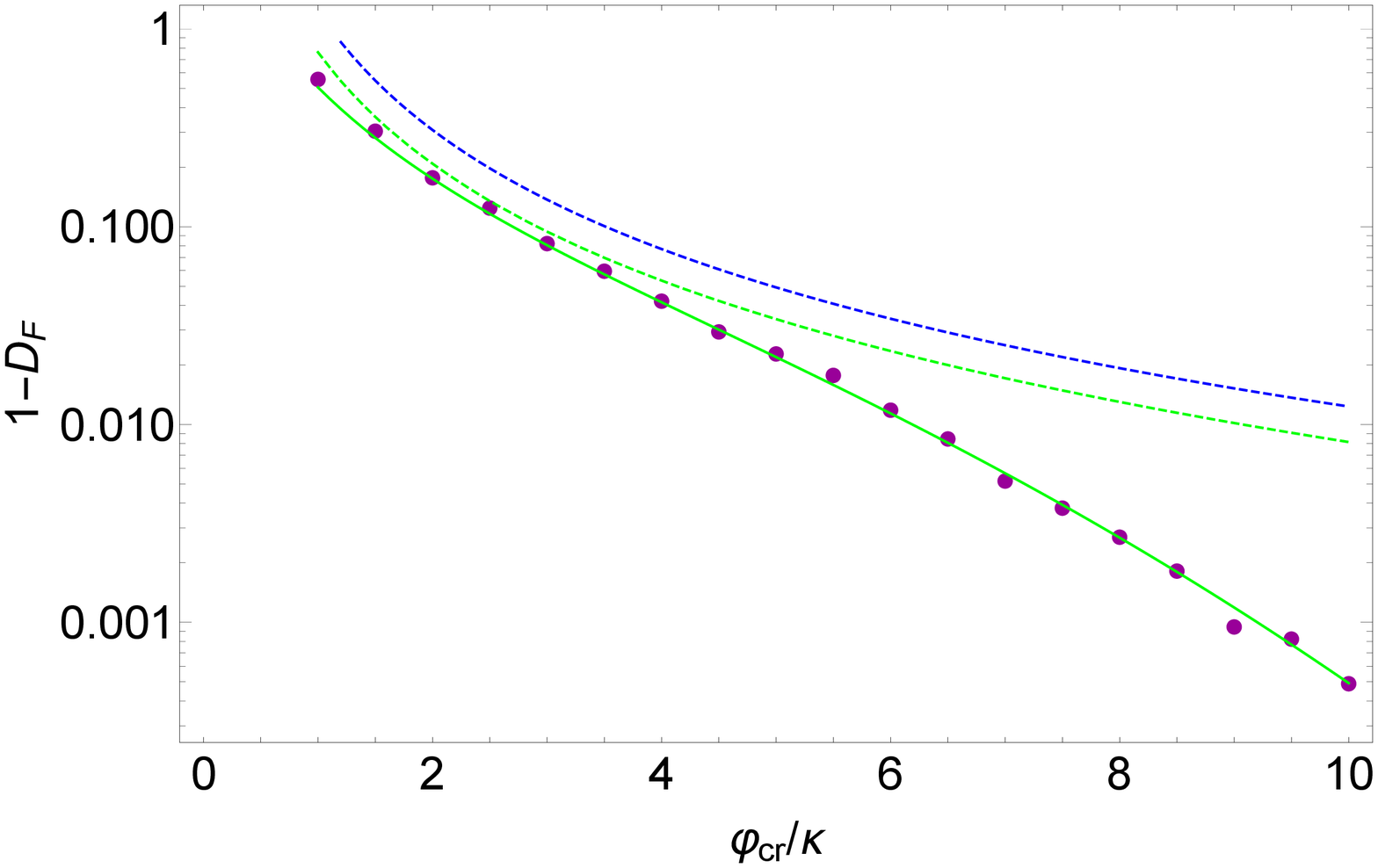}
\caption{\label{fig:logfractaldimension} Change in fractal dimension $1 - D_{F}$ vs $\varphi_{cr}/\kappa$. Top panel is network-I and bottom panel is network-II. Dark red (magenta) dots are simulation data up to 18 2-foldings, the two green curves are from kernel evolution with $\epsilon = \ln 2$ for the solid; while $\epsilon = \ln 1.1$ and steady state for the dashed. The dashed blue curve is the analytical result (eq.~\eqref{eq:fractaldimension}).}
\label{FractalDimension}
\end{figure}

First, we compare fractal dimensions. Fig.~\ref{FractalDimension} shows the comparison between analytics and simulation for network-I (upper panel) and network-II (lower panel). Note that since we have only gone up until $18$ 2-foldings in this figure, we have only converged to the steady state behavior for $\varphi_{cr}/\kappa$ around $\sim 3.5$, or so. The interpolation between discrete and continuous regime also suggests that the theory calculation gives the right behavior.

\begin{figure}[t]
\centering
\includegraphics[width=1\columnwidth]{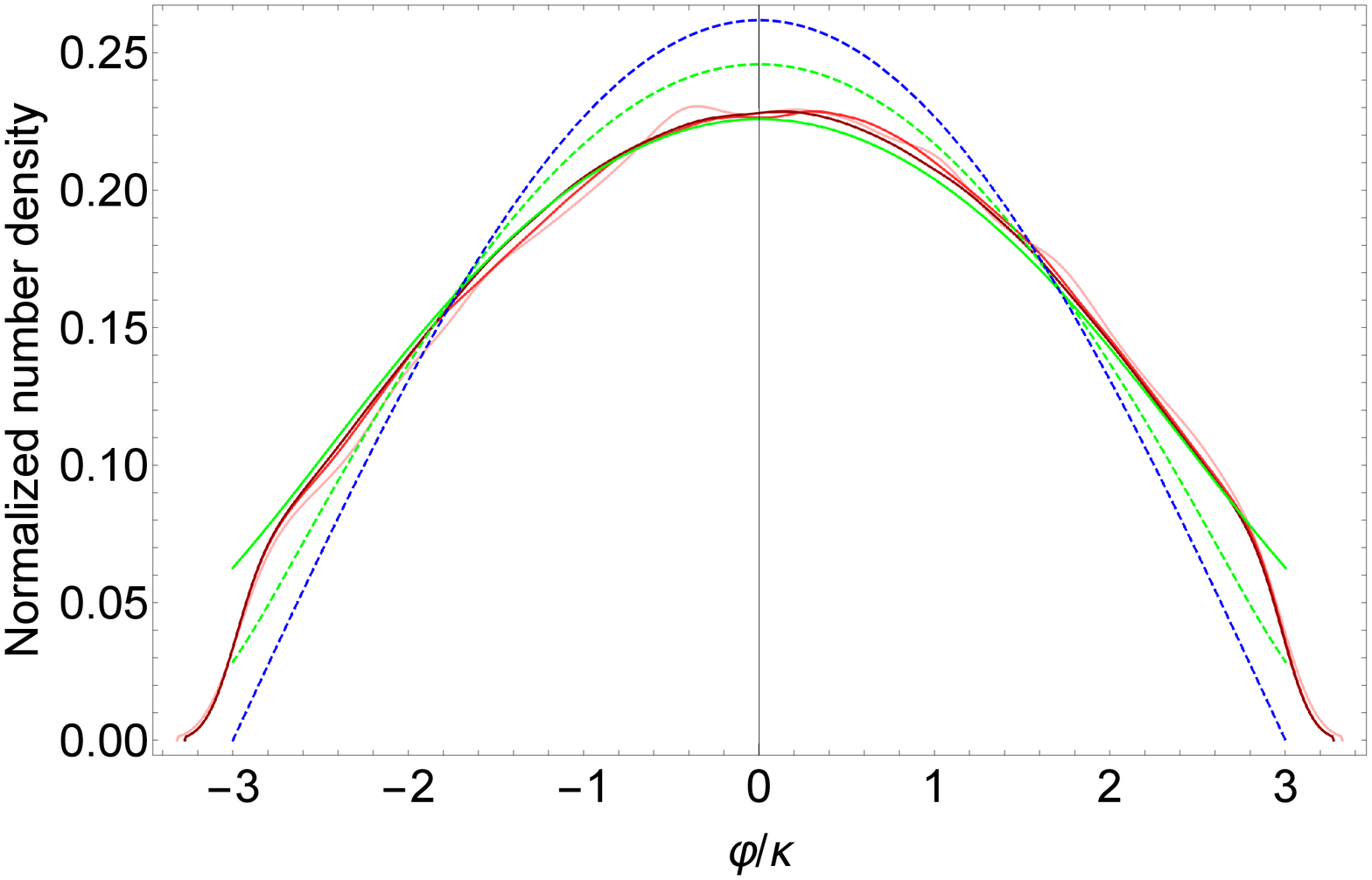}\vspace{\sz}
\includegraphics[width=1\columnwidth]{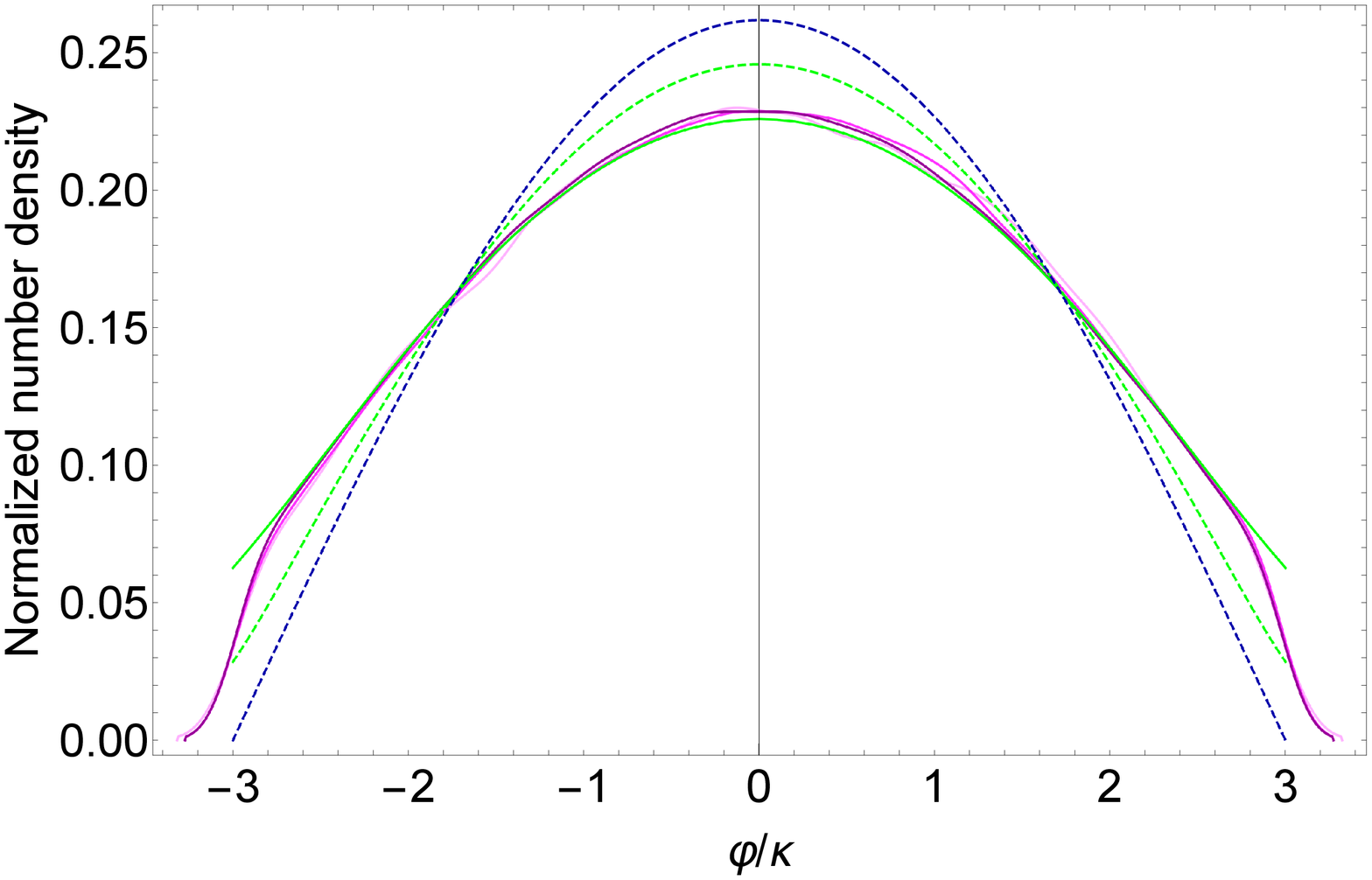}
\caption{Number density $\tilde{\rho}$ vs crunch value $\varphi/\kappa$ for $\varphi_{cr}/\kappa = 3$. Top panel is network-I and bottom panel is network-II. All the red (magenta) curves in increasing darkness represent simulation runs for $12, 15$, and $18$ $2$-foldings respectively. The solid green curve is from kernel propagation with $\epsilon = \ln 2$ at steady state, while the dashed green is for $\epsilon = \ln 1.1$ at steady state. Dashed blue is the analytical result eq.~\eqref{eq:steady.state.distribution}.}
\label{NumberDensity3p5}\end{figure}

\begin{figure}[t]
\centering
\includegraphics[width=1\columnwidth]{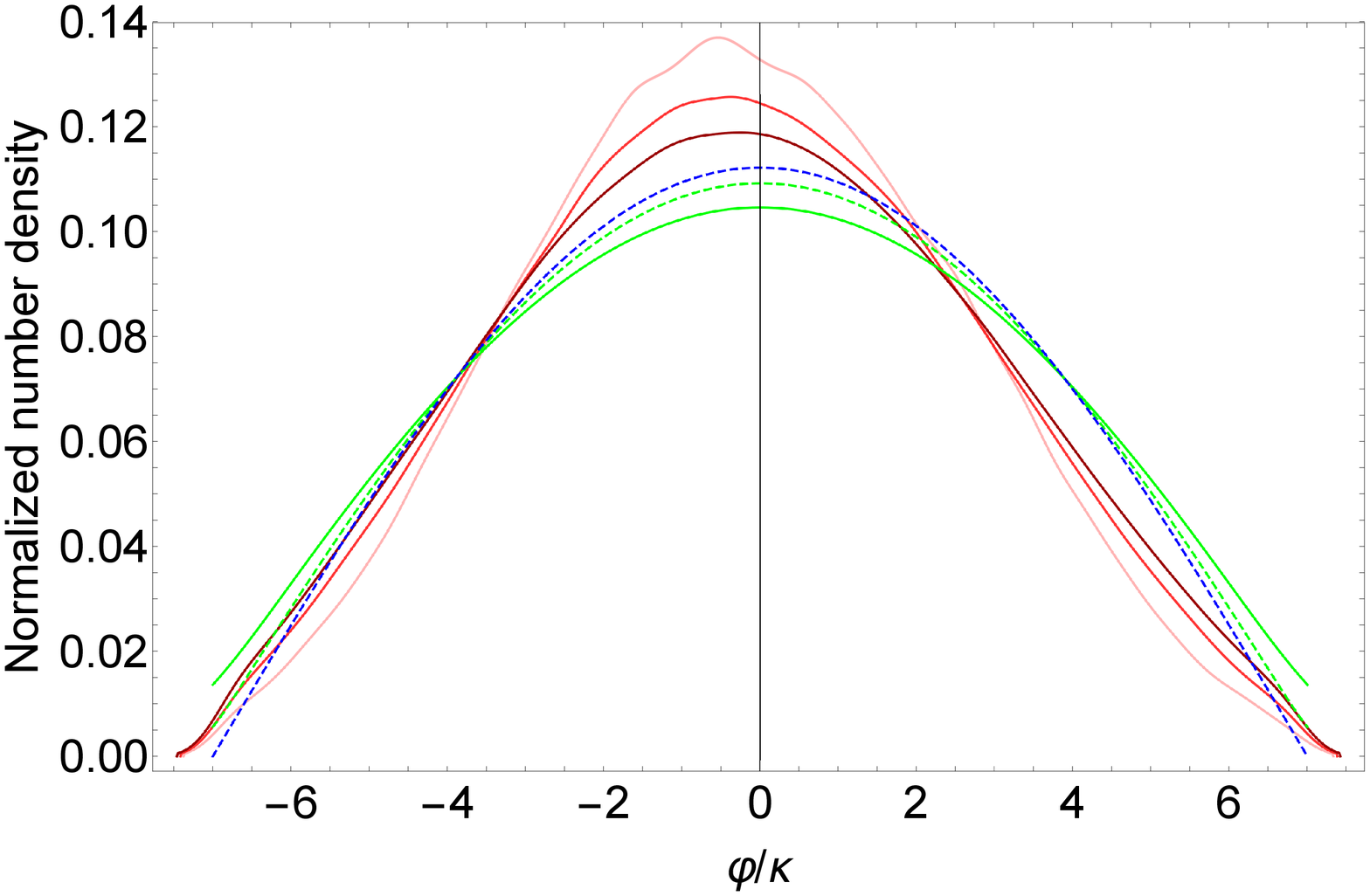}\vspace{\sz}
\includegraphics[width=1\columnwidth]{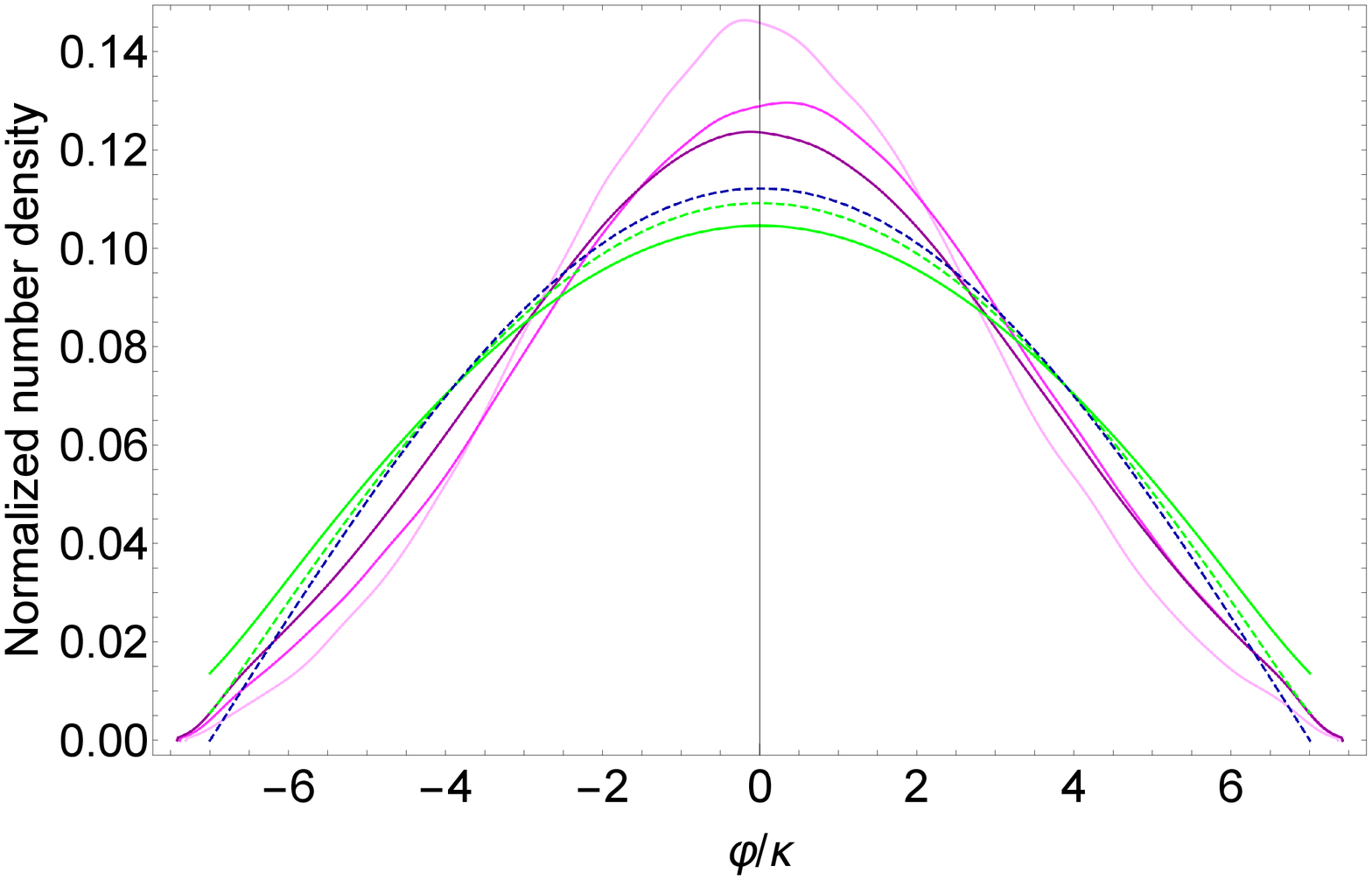}
\caption{Number density $\tilde{\rho}$ vs crunch value $\varphi/\kappa$ for $\varphi_{cr}/\kappa = 7$. Top panel is network-I and bottom panel is network-II. The convention is the same as in Fig.~\ref{NumberDensity3p5}. Here it has yet to properly converge to the steady state distribution.}
\label{NumberDensity7}
\end{figure}

Next, we present plots for the normalized number density $\tilde{\rho}$, for the two networks. Fig.~\ref{NumberDensity3p5} is for $\varphi_{cr}/\kappa = 3$ and Fig.~\ref{NumberDensity7} is for $\varphi_{cr}/\kappa = 7$. The curves are for $12, 15$, and $18$ $2$-foldings. The aforementioned convergence is evident from these plots.

\begin{figure}[t]
\centering
\includegraphics[width=1\columnwidth]{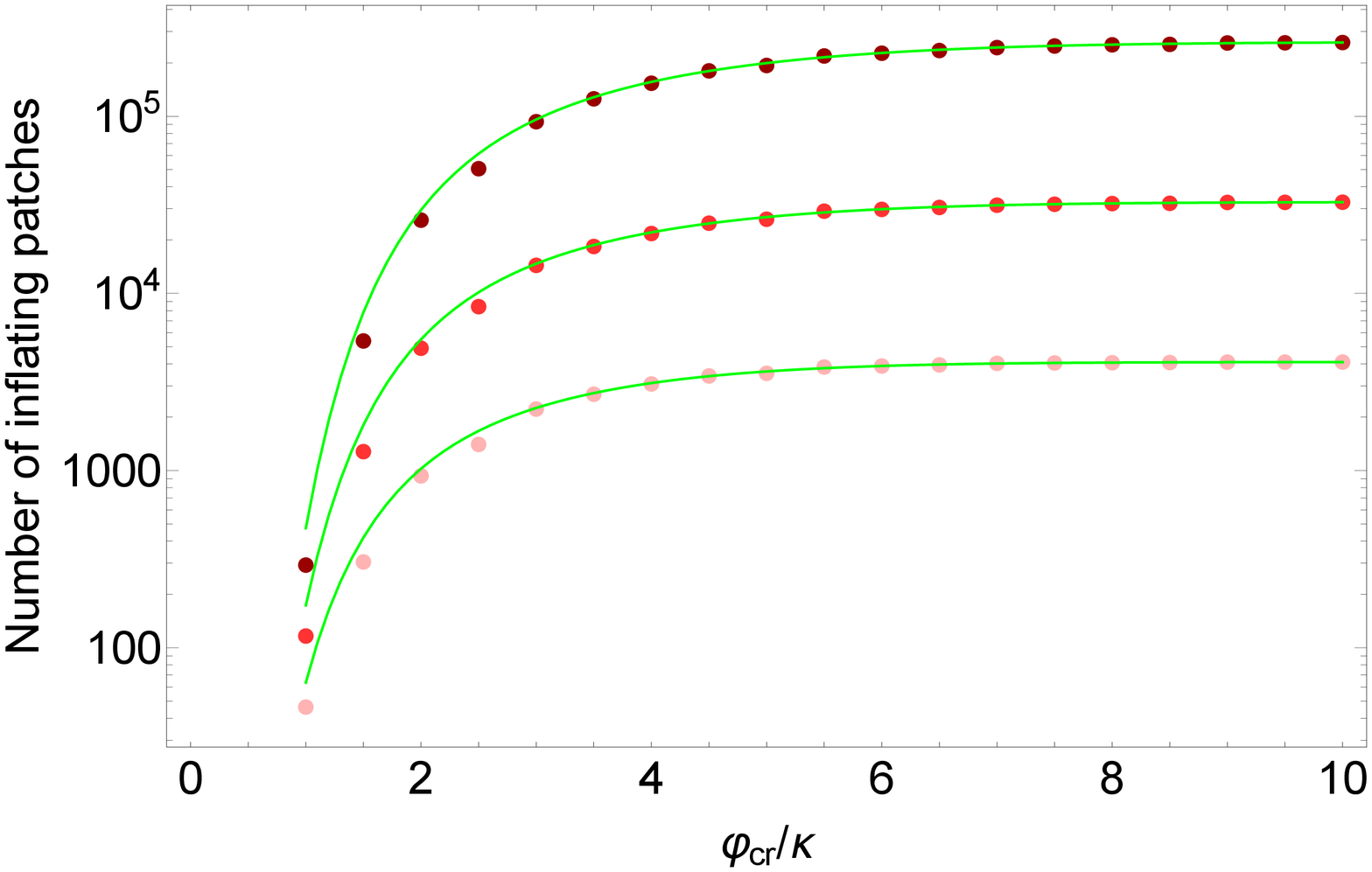}\vspace{\sz}
\includegraphics[width=1\columnwidth]{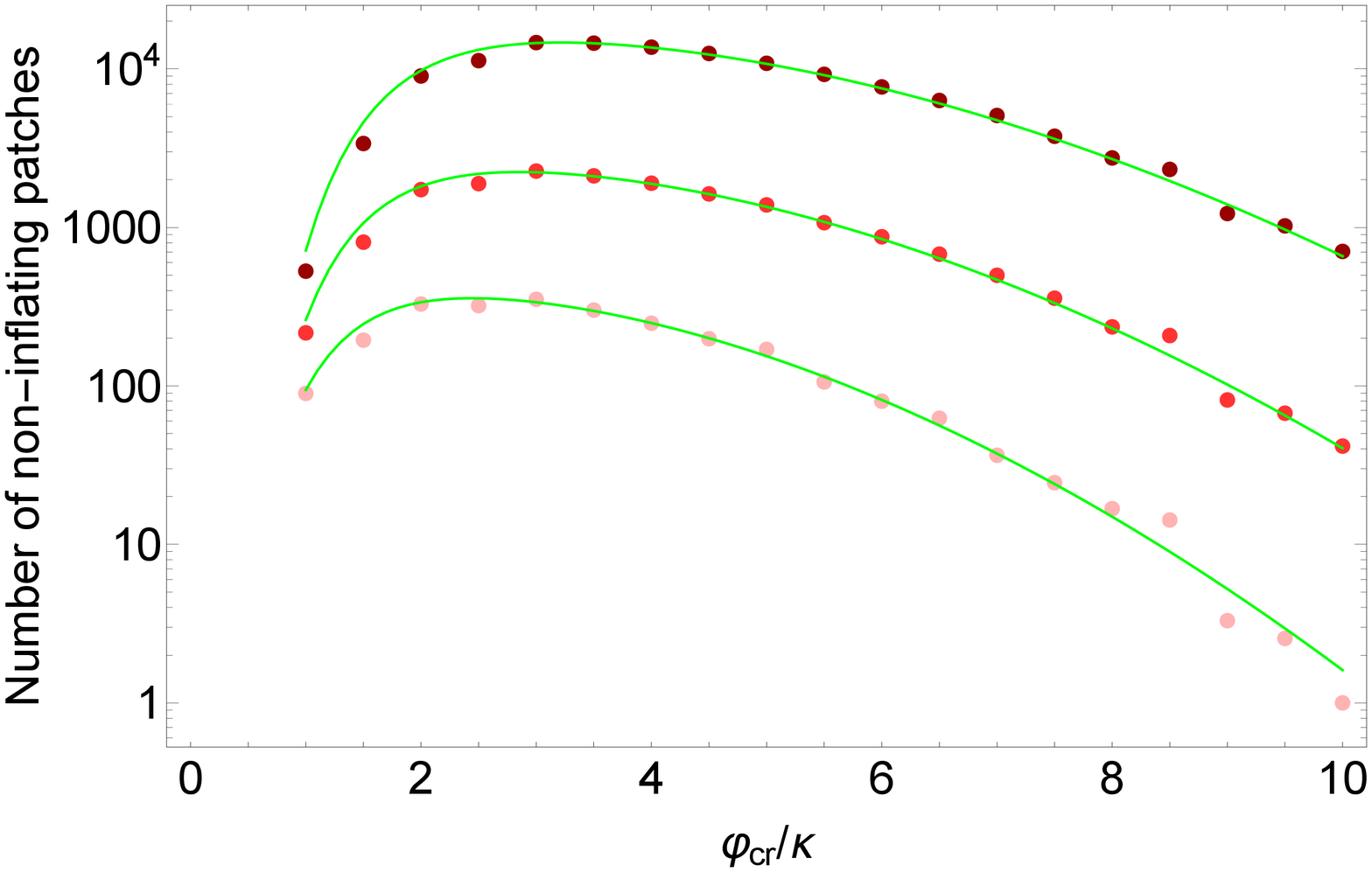}
\caption{Top panel: Number of inflating Hubble patches vs $\varphi_{cr}/\kappa$ for $12, 15$, and $18$ $2$-foldings. Bottom panel: Number of thermalized/crunched patches vs $\varphi_{cr}/\kappa$ for $12, 15$, and $18$ $2$-foldings. Both plots are for network-I. In network-II there is an unphysical exponential growth in the size of the thermalized/crunch regions, so it is of little interest to plot here. Dots (in increasing order of redness) represent simulation data while the corresponding green curves represent kernel evolution with $\epsilon = \ln 2$ and the corresponding number of 2-foldings.}
\label{NumberInflatingCrunched}\end{figure}

In Fig.~\ref{NumberInflatingCrunched} we show the comparison between the number of inflating Hubble patches in network-I and the kernel propagation method. This shows excellent agreement. Furthermore, we also show the comparison between the number of thermalized/crunched Hubble patches, and we see excellent agreement also. For network-II the number of inflating patches is very similar. However, the number of thermalized/crunches Hubble patches is artificially enhanced by the method, and so this particular quantity is not of interest to report on here.

Having shown these various results, which provide evidence in support of the simulation networks, and points towards steady state behavior, we can now move on to the discussion of the distribution of sizes of inflating regions within this eternally inflating Universe.

\section{Statistics of inflating regions}

Before reporting on our simulation results and understanding, we would like to point out that there are two {\em naive} arguments regarding typical sizes of inflating regions $\langle l \rangle$. Both lead to the expectation that the typical distance is exponentially large in some power of the crunch value $\varphi_{cr}$. In the first argument, one begins by saying that the typical number of e-foldings it takes for the field to get to get from its starting value $\varphi=0$ to its crunch value $\varphi=\pm\varphi_{cr}$, is of the order of $N_{typ}\sim\varphi_{cr}^2/\kappa^2$ owing to its simple random walk. But by this time, the Universe has expanded by $e^{N_{typ}}$ and thus the typical distance between non-inflating regions (or equivalently, the typical size of inflating regions) is $\sim e^{c_1\varphi_{cr}^2/\kappa^2}$, where $c_1=\mathcal{O}(1)$ prefactor. 

In the second argument, one tries to improve on this by arguing that in order to estimate typical sizes, we must make sure that we have a clustered set of Hubble patches (or a connected chain in the simple case of 1 dimensions). With the random walk behavior of the field, a naive calculation for the probability for a single Hubble patch to stay within $(-\varphi_{cr},\varphi_{cr})$ gives $\text{Erf}\left(\varphi_{cr}/(\sqrt{2N}\kappa)\right)$. Treating each Hubble patch independently, one then raises this probability to $e^{DN}$ and demands that this quantity be an order 1 number. This calculation gives $N_{typ} \sim \varphi_{cr}/\kappa$ and therefore the typical sizes of inflating regions is $\sim e^{c_2\varphi_{cr}/\kappa}$, where $c_2=\mathcal{O}(1)$ prefactor. 

However, both of these naive pictures in fact misses a crucial aspect of the physics. While the second school of thought tries to improve on the first one by demanding a clustered region of inflating Hubble patches, both of them completely ignore the steady state behavior of this eternally inflating Universe. Since the fractions of inflating and non-inflating patches go to constants, this means that the probabilities to find either of them become constants too. After noting this crucial fact, if one then treats each of the inflating Hubble patches as being roughly independent, one finds a very different answer. The probability that a patch is inflating is simply equal to the fraction $f$. In 1 spatial dimension, given an inflating Hubble patch, the probability that there is a clustered region of $l$ such patches is $\approx f^{l - 1}$ (modulo an overall normalization) giving an exponential distribution in sizes $l$ (in units of Hubble $H^{-1}$):
\begin{equation}
P_{l} \approx \dfrac{f^{(l-1)}}{\sum^{\infty}_{l=1}f^{(l-1)}} = \left(1-f\right) f^{l-1},
\end{equation}
with a corresponding average size
\begin{equation}
\langle l\rangle\approx (1-f)\sum^{\infty}_{l=1}l\,f^{(l-1)} = \left(\dfrac{1}{1-f}\right),
\end{equation}
which for large $\varphi_{cr}$ goes to the following (using eq.~\eqref{eq:fractions} with $D=1$ and $\kappa=1/\sqrt{2\pi}$)
\begin{equation}
\langle l\rangle \approx {16\,\varphi_{cr}^2\over\pi}.
\label{lpredict}\end{equation}
So the growth with $\varphi_{cr}$ is only a mere power law, rather than exponential. Furthermore, in $D$ spatial dimensions, the generalization of this reasoning is to the power law $\langle l \rangle\propto\varphi_{cr}^{2/D}$. So for higher $D$, the growth with $\varphi_{cr}$ is expected to be even slower.

\begin{figure}[t]
\centering
\includegraphics[width=1\columnwidth]{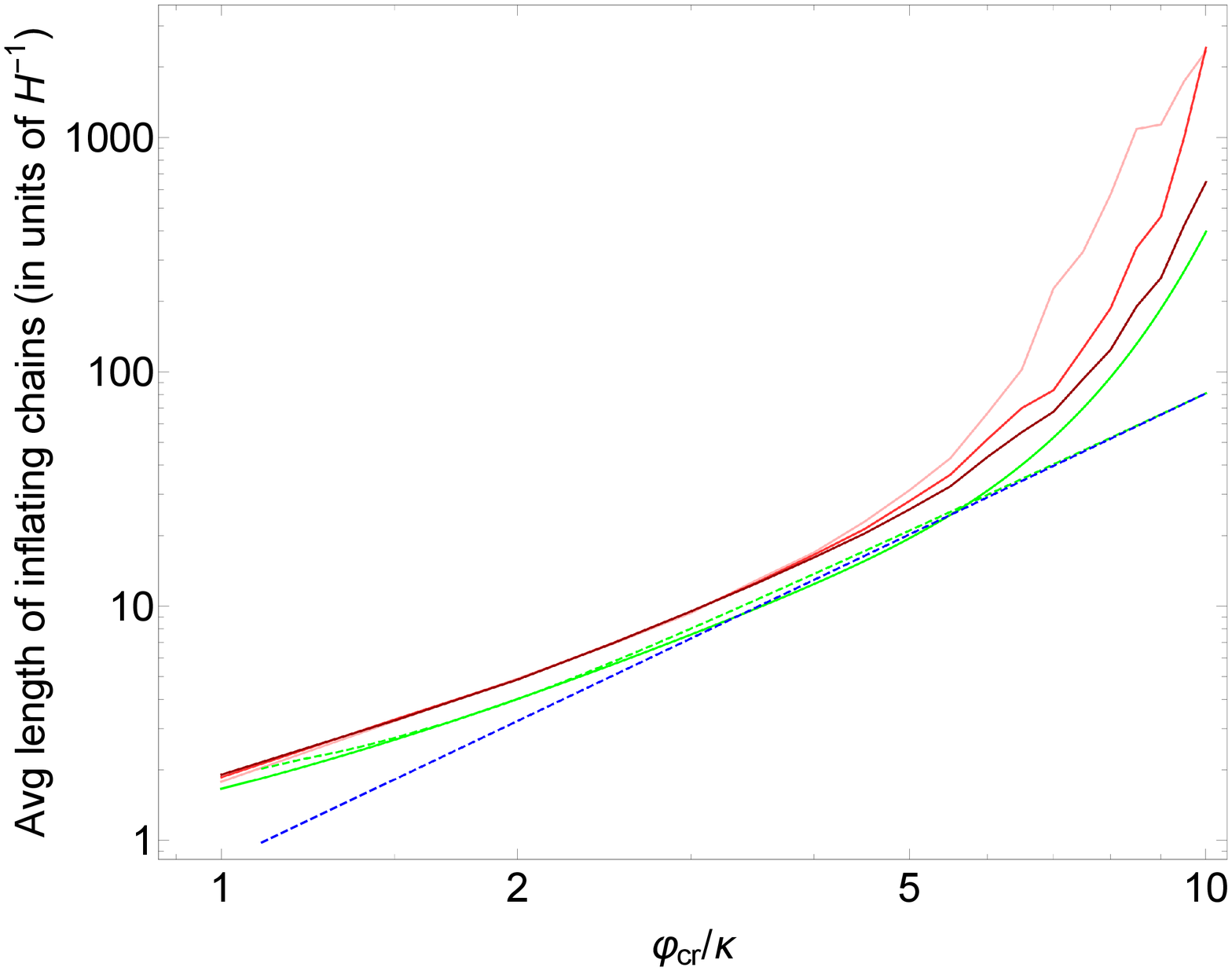}\vspace{\sz}
\includegraphics[width=1\columnwidth]{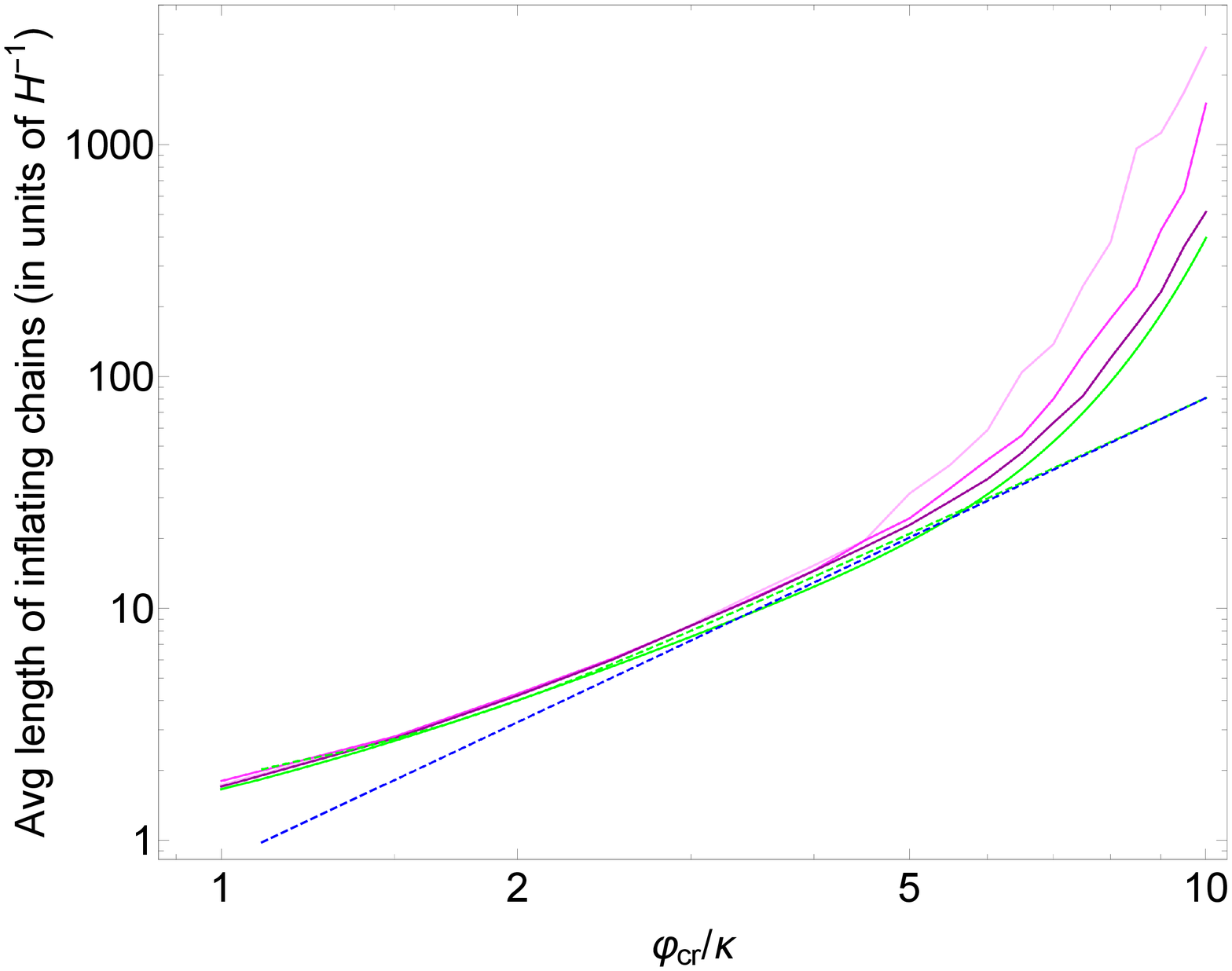}
\caption{Average distances $\langle l\rangle$ vs $\varphi_{cr}/\kappa$. Top panel is network-I and bottom panel is network-II. Red (magenta) curves with increasing darkness are for $12$, $15$, and $18$ 2-foldings respectively. The solid and dashed green curves are from kernel propagation with $\epsilon = \ln 2$ only up-to $18$ 2 foldings, and $\epsilon = \ln 1.1$ at steady state respectively. The dashed blue curve is the theoretical expectation in steady state for high $\varphi_{cr}/\kappa$ eq.~\eqref{lpredict}. Note that this is on a log-log scale, so that as curves asymptote to a straight line, it implies power law dependence on $\varphi_{cr}$.} 
\label{AverageDistances}
\end{figure}

Fig.~\ref{AverageDistances} shows the trend of average sizes of inflating regions as we increase the number of 2-foldings. The converging trend of simulation red (magenta) curves towards a linear curve, is in agreement with our estimates (note that this is on a log-log scale). Again we note that we have converged to the steady state behavior only up until around $\varphi_{cr}/\kappa \sim 3.5$ even for the maximum number of 2-foldings shown. The upward trend is consistent with the theoretical expectation (given by the green curve) and is anticipated to approach a straight line (on this log-log scale) as we increase the number of 2-foldings.

\begin{figure}[t]
\centering
\includegraphics[width=1\columnwidth]{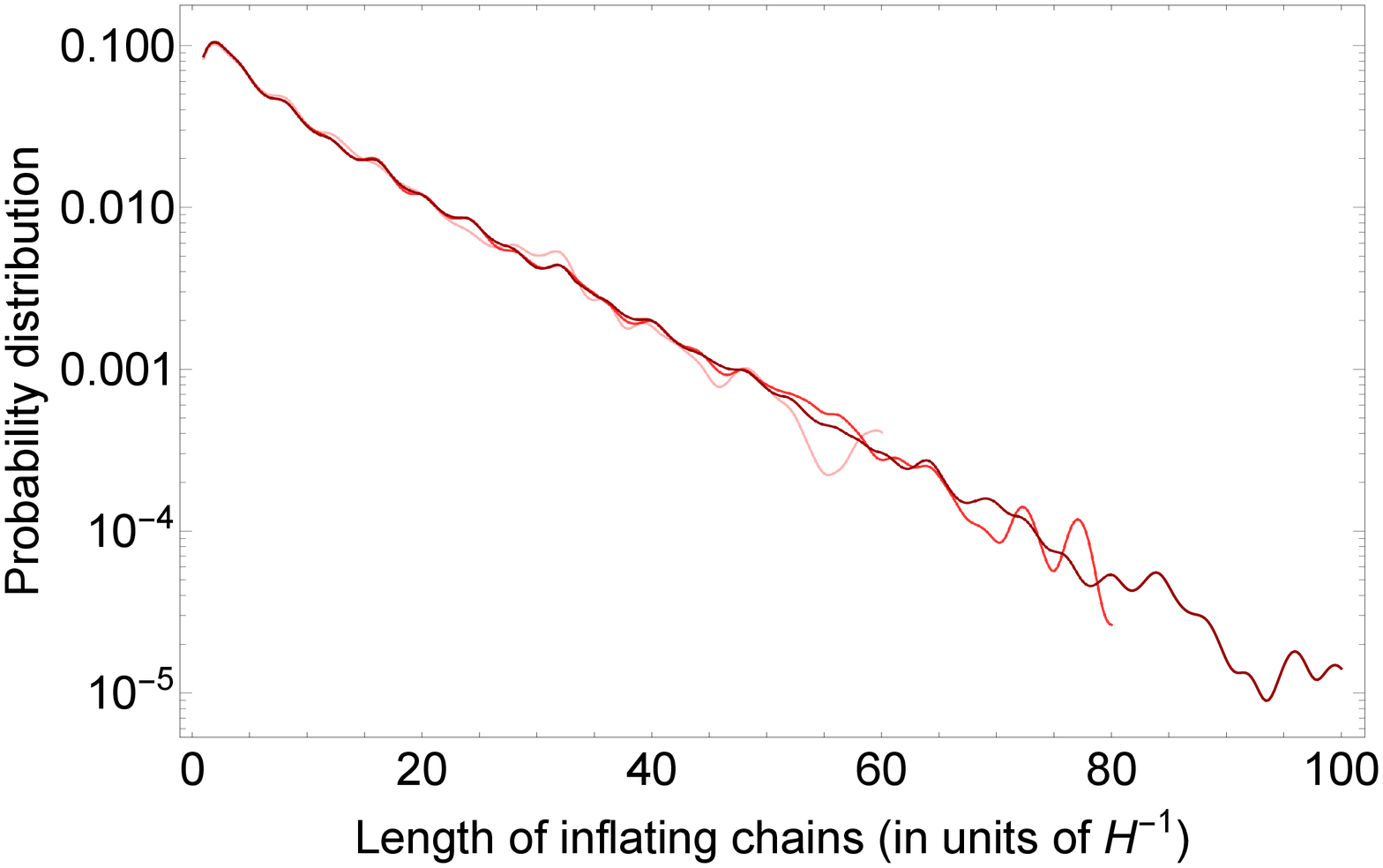}\vspace{\sz}
\includegraphics[width=1\columnwidth]{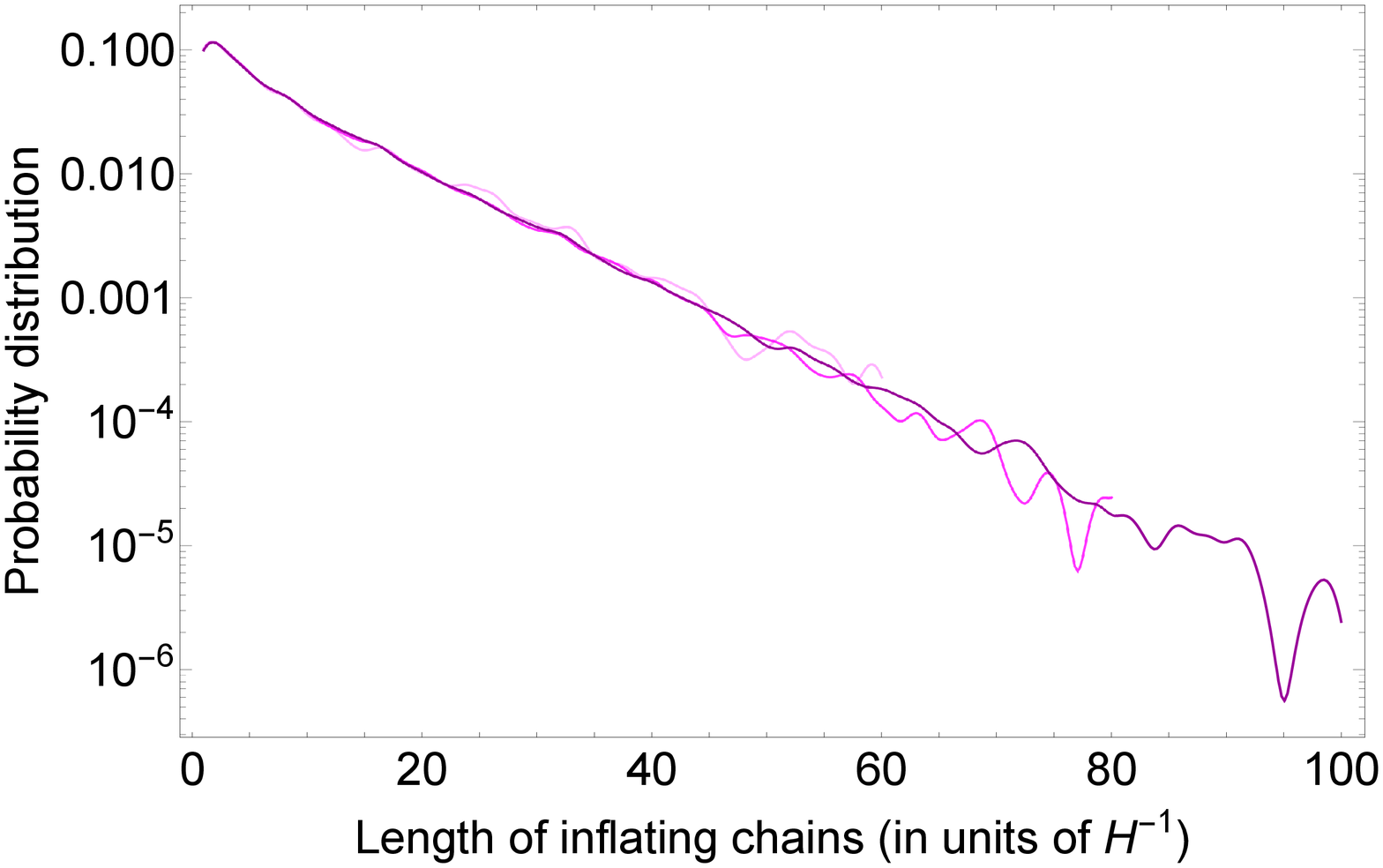}
\caption{Distribution in lengths for inflating regions for $\varphi_{cr}/\kappa = 3$ for $12$, $15$, and $18$ 2-foldings (in increasing darkness). Top panel is network-I and bottom panel is network-II.} 
\label{Distribution3}
\end{figure}

\begin{figure}[t!]
\centering
\includegraphics[width=1\columnwidth]{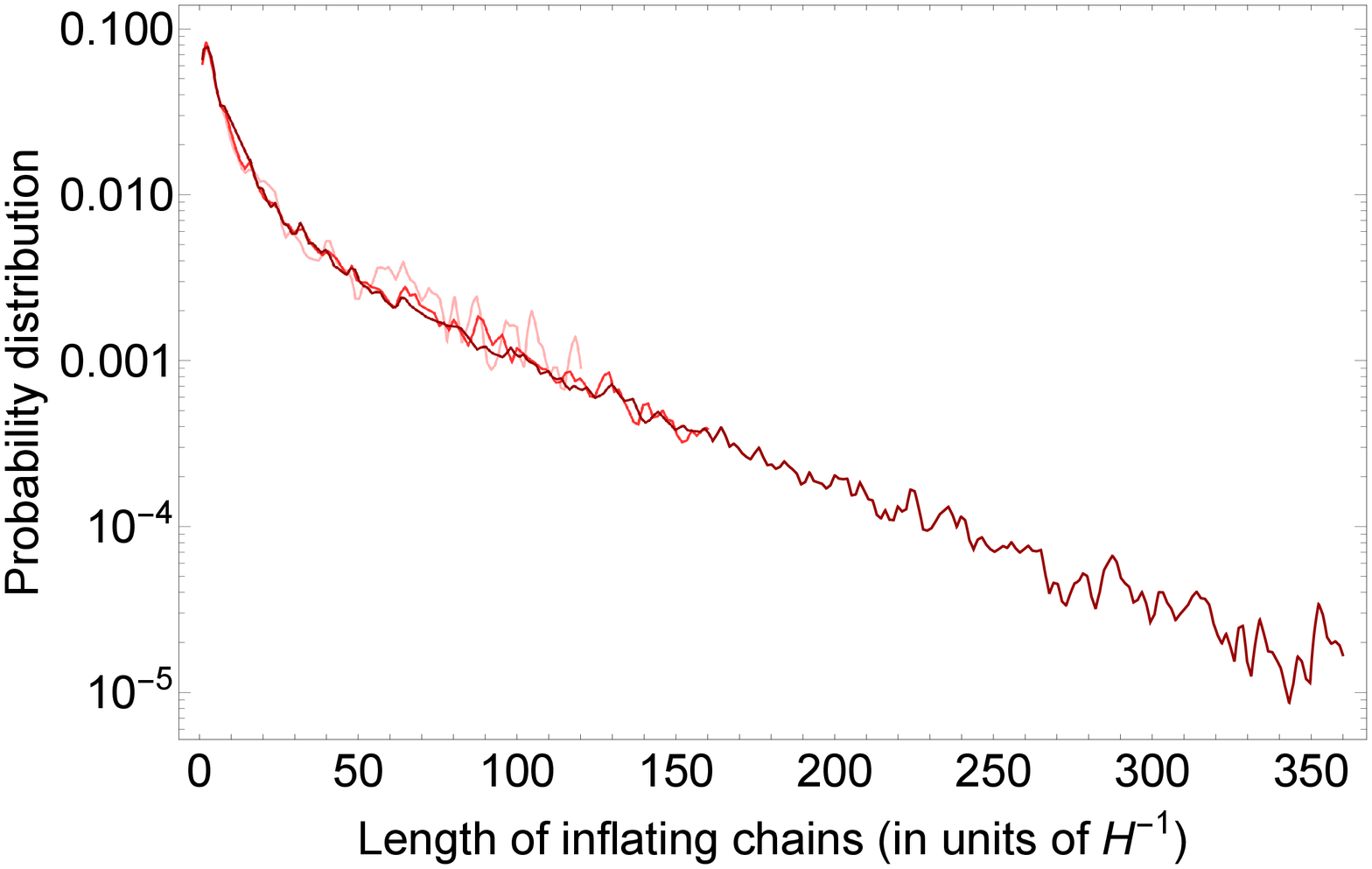}\vspace{\sz}
\includegraphics[width=1\columnwidth]{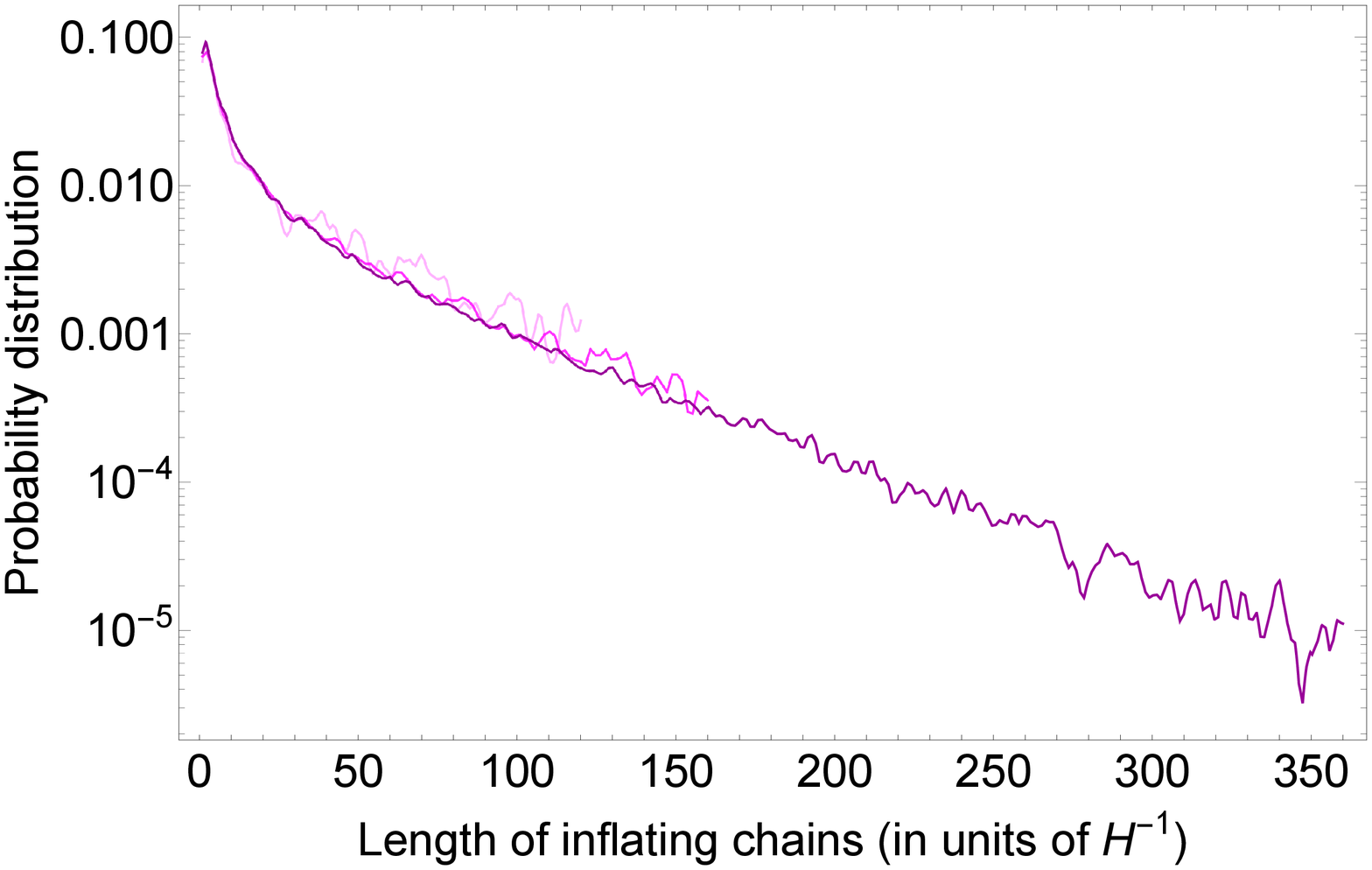}
\caption{Distribution in lengths for inflating regions for $\varphi_{cr}/\kappa = 5$ for $12$, $15$, and $18$ 2-foldings (in increasing darkness). Top panel is network-I and bottom panel is network-II.} 
\label{Distribution5}
\end{figure}

Next, for the distribution of sizes of inflating regions, we do indeed find indication of a decaying exponential behavior, at least for sufficiently large sizes. This is seen clearly in Fig.~\ref{Distribution3} with $\varphi_{cr}/\kappa=3$ (note the log-linear scale). For small to reasonably large sizes, however, the fall off appears to be only a power law; so there is still a non-trivial chance to obtain regions with size appreciably larger than the average. This is due to correlations among nearby Hubble patches that is not captured in the above simplistic analysis. This can be seen more clearly in Fig.~\ref{Distribution5} with $\varphi_{cr}/\kappa = 5$ out to $l\sim50$ or so. However, at larger distances, we again see exponential suppression.


\section{Statistics of non-inflating regions}

\begin{figure}[t]
\centering
\includegraphics[width=1\columnwidth]{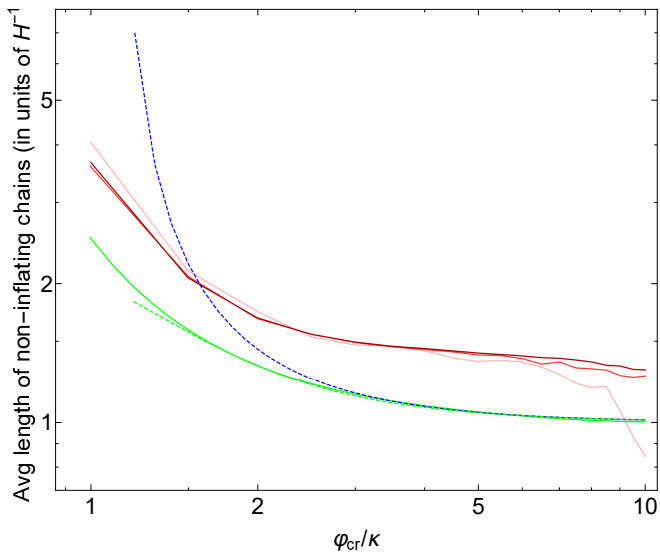}\vspace{\sz}
\caption{Average distances $\langle \bar{l}\rangle$ vs $\varphi_{cr}/\kappa$ from network-I. Red curves with increasing darkness are for $12$, $15$, and $18$ 2-foldings respectively. The solid and dashed green curves are from kernel propagation with $\epsilon = \ln 2$ only up-to $18$ 2 foldings, and $\epsilon = \ln 1.1$ at steady state respectively. The dashed blue curve is the theoretical expectation in steady state for high $\varphi_{cr}/\kappa$ eq.~\eqref{lbarpredict}.} 
\label{AverageDistancescrunched}
\end{figure}

\begin{figure}[t]
\centering
\includegraphics[width=1\columnwidth]{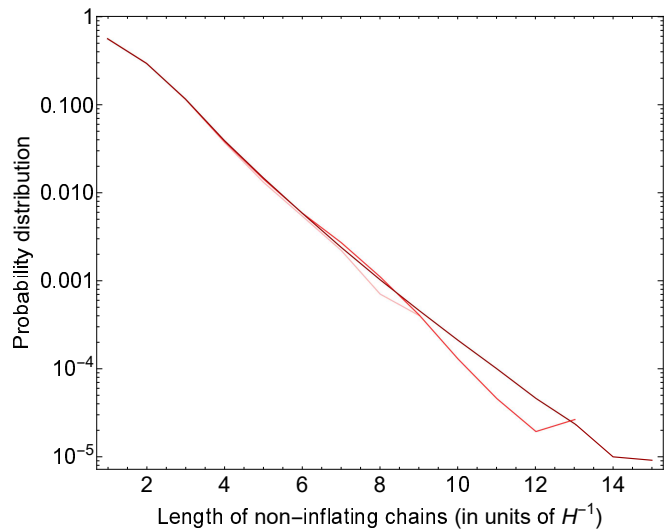}\vspace{\sz}
\includegraphics[width=1\columnwidth]{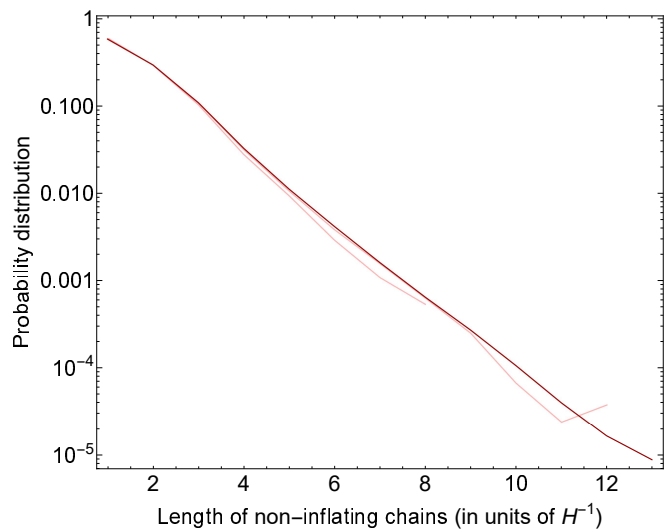}
\caption{Distribution in lengths for non-inflating regions for $12$, $15$, and $18$ 2-foldings (in increasing darkness), from network-I. Top panel is for $\varphi_{cr}/\kappa = 3$ and bottom panel is for $\varphi_{cr}/\kappa=5$.} 
\label{Distributioncrunched}
\end{figure}

For the sizes of non-inflating regions, we again find a power law behavior as suggested by the steady state combinatorics. The fraction of non-inflating regions is $\bar{f}$ giving the following exponentially decaying distribution in sizes $\bar{l}$ (in units of $H^{-1}$)
\begin{equation}
P_{\bar{l}} \approx \dfrac{\bar{f}^{(\bar{l}-1)}}{\sum^{\infty}_{\bar{l}=1}\bar{f}^{(\bar{l}-1)}} = \left(1-\bar{f}\right) \bar{f}^{\bar{l}-1},
\end{equation}
with the corresponding average size
\begin{equation}
\langle \bar{l}\rangle \approx (1-\bar{f})\sum^{\infty}_{\bar{l}=1}\bar{l}\,\bar{f}^{(\bar{l}-1)} = \left(\dfrac{1}{1-\bar{f}}\right),
\end{equation}
which for large $\varphi_{cr}$ goes to the following (using eq.~\eqref{eq:fractions} with $D=1$ and $\kappa=1/\sqrt{2\pi}$)
\begin{equation}
\langle \bar{l}\rangle \approx 1 + \dfrac{\pi}{16\,\varphi_{cr}^2}.
\label{lbarpredict}\end{equation}
Similarly like before in $D$ spatial dimensions, the generalization of this reasoning is to the power law $\langle \bar{l} \rangle \sim (1+\mathcal{O}(1)\kappa^2/\varphi_{cr}^2)^{1/D}$. Fig.~\ref{AverageDistancescrunched} shows the trend of average sizes of non-inflating regions as we increase the number of 2-foldings, and fig. ~\ref{Distributioncrunched} show the distributions in sizes for $\varphi_{cr}/\kappa=3$ and $5$ respectively. 

\section{Summary and Discussion}

To summarize, we have found that in a simple illustrative scenario of eternal inflation (with $H = const$) in the presence of a scalar field $\varphi$ with a constant potential $V(\varphi)$ within $(-\varphi_{cr},\varphi_{cr})$, the average size of inflating regions only grows as a power law in $\varphi_{cr}$, and the distributions in sizes falls off exponentially for sufficiently large distances. This result is quite different from a naive understanding of general eternal inflation scenarios, where one might expect average sizes of inflating regions to have some sort of exponential dependence on $\varphi_{cr}$. We have found this not to be the case since this eternally inflating Universe ultimately gets to an equilibrium situation where the fractions of inflating (non-inflating) Hubble patches become constant(s), and only grow to unity (dying out to zero) power law fast in $\varphi_{cr}$. However the distribution does have some large width to it due to correlations among nearby Hubble patches, so there is still a non-trivial probability to create regions that are much larger than the average. Similarly for non-inflating regions we have found that the distribution is an exponential decay law with typical sizes decaying like a power law in $1/\varphi_{cr}$.

We have also established a simple kernel/integral evolution technique that not only serves to provide interpolation between discrete simulation and continuous theory, but is also useful for investigating more interesting and involved scenarios where the potential of the scalar field has some slope, resulting in a dominant slow roll phase making analytical calculations formidable in general. We leave more interesting cases of $V(\varphi) \neq const$ and time varying $H$ for future work, along with an application of our results to the case of SM Higgs in the early Universe. In the latter case the multi-component nature of the Higgs doublet can play an important role in its statistics \cite{Hertzberg:2018kyi}.

Improved analytical estimates can be made in future work. The (relatively small) mismatch between analytics and simulation for the average sizes of inflating regions (seen here in Fig.~\ref{AverageDistances} and \ref{AverageDistancescrunched}) seems to be due to two reasons. Firstly, we are comparing with discrete simulation which requires choosing the right cut-off for the theory. Secondly, and perhaps more importantly, treating each Hubble patch independently in the simplified combinatorics is obviously not precise since nearby patches have some correlation, and so there is ambiguity in how many Hubble patches to treat as one independent entity. What is important however is the power law dependence of average sizes over $\phi_{cr}$, which indeed is in agreement.

In order to understand the nature of distributions even better, we have also looked at several statistical moments in order to check if they grow with increasing number of 2-foldings or if they have some power law behavior in their tails. We did not find any such behavior for both inflating and non-inflating sizes. In the Appendix we provide some plots for these higher moments (for network-I).

Although the \textit{average} sizes of inflating (and non-inflating) regions is only power law in $\varphi$, rather than exponential, it should be noted that this by no means undermines the general idea of inflation which is to provide a huge homogeneous/isotropic Universe. The growth of higher central moments with $\varphi_{cr}$ indicates a non-trivial chance of obtaining much larger regions. Also, what we have studied here is that of a flat potential for a field that falls off with a ``cliff" to sudden thermalization/crunching; this is a universe that is completely dominated by quantum diffusion. The introduction of a tilted potential that allows for a prolonged phase of classical domination over quantum diffusion can lead to the production of much larger (thermalized) homogeneous regions. The corresponding distribution of sizes is anticipated to be rather different and will be the subject of future work \cite{Forthcoming}.

\section*{Acknowledgments}

We would like to thank Matt Kleban, Andrei Linde, Shao-Jiang Wang, and especially Alex Vilenkin for useful discussions. M. P. H. is supported in part by National Science Foundation Grant No. PHY-1720332, a Gordon Godrey fellowship, and a JSPS fellowship.

\appendix

\section{Higher moments for distribution in size of regions}

\begin{figure}[t]
\centering
\includegraphics[width=0.95\columnwidth]{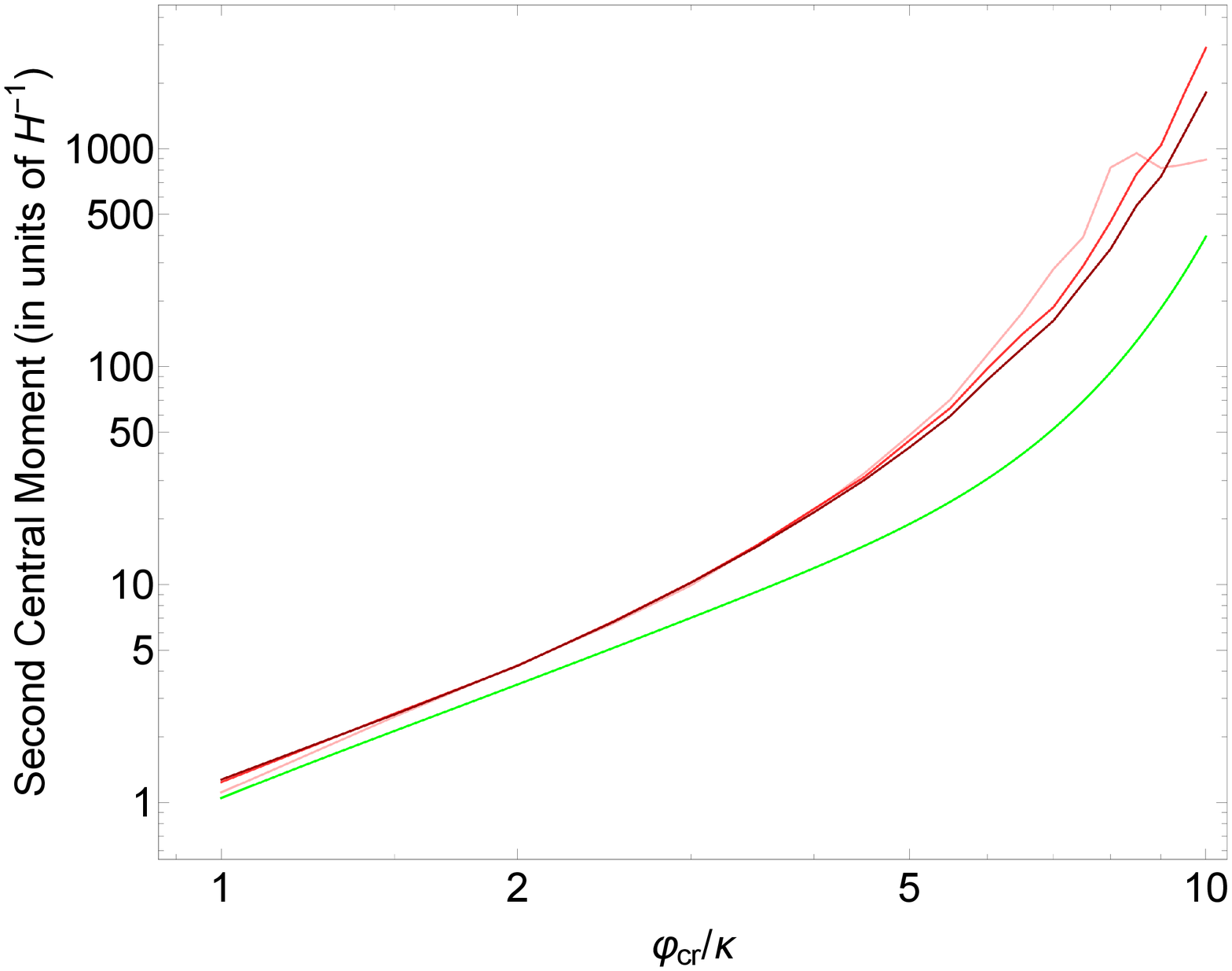}\vspace{\sz}
\includegraphics[width=0.95\columnwidth]{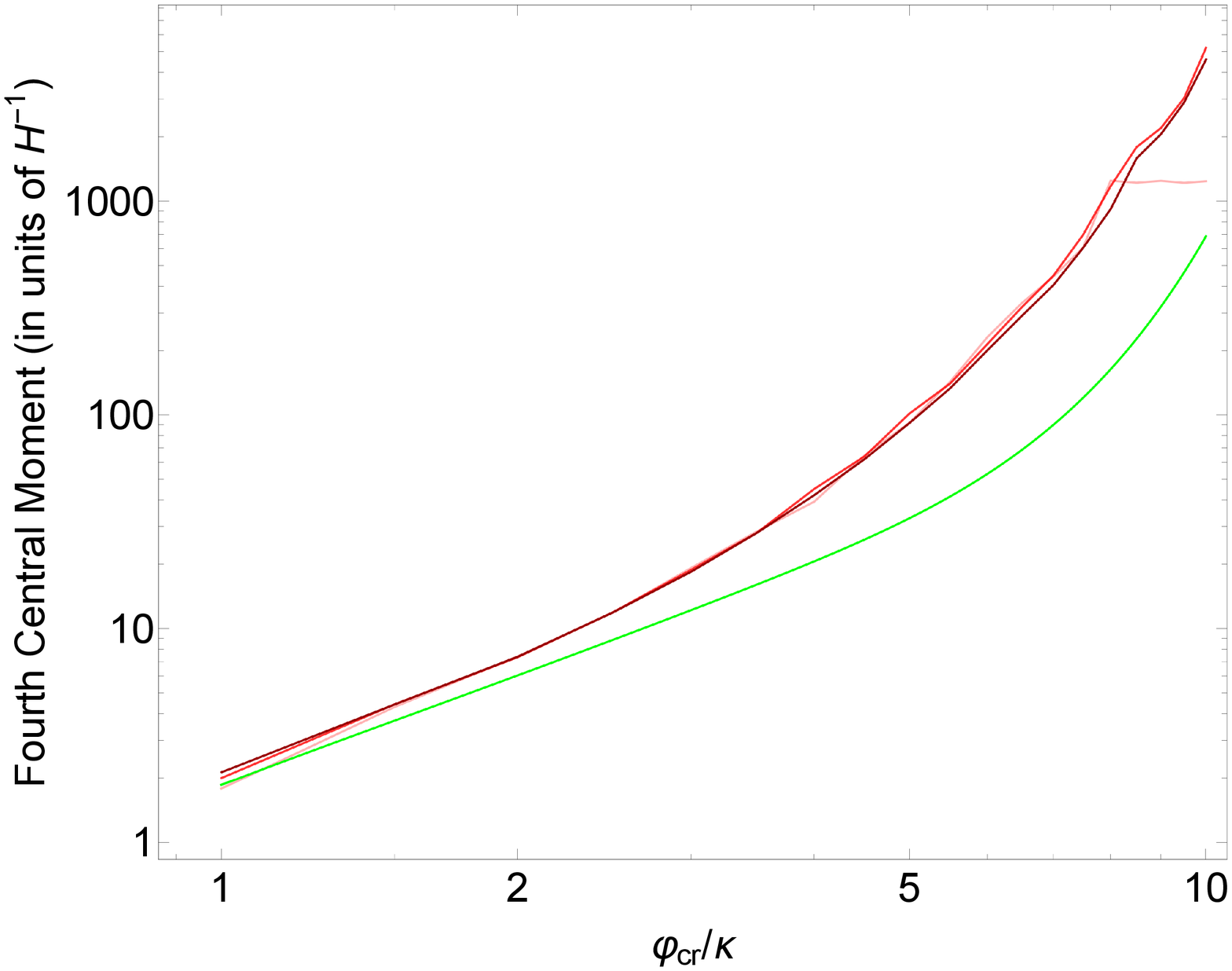}\vspace{\sz}
\includegraphics[width=0.95\columnwidth]{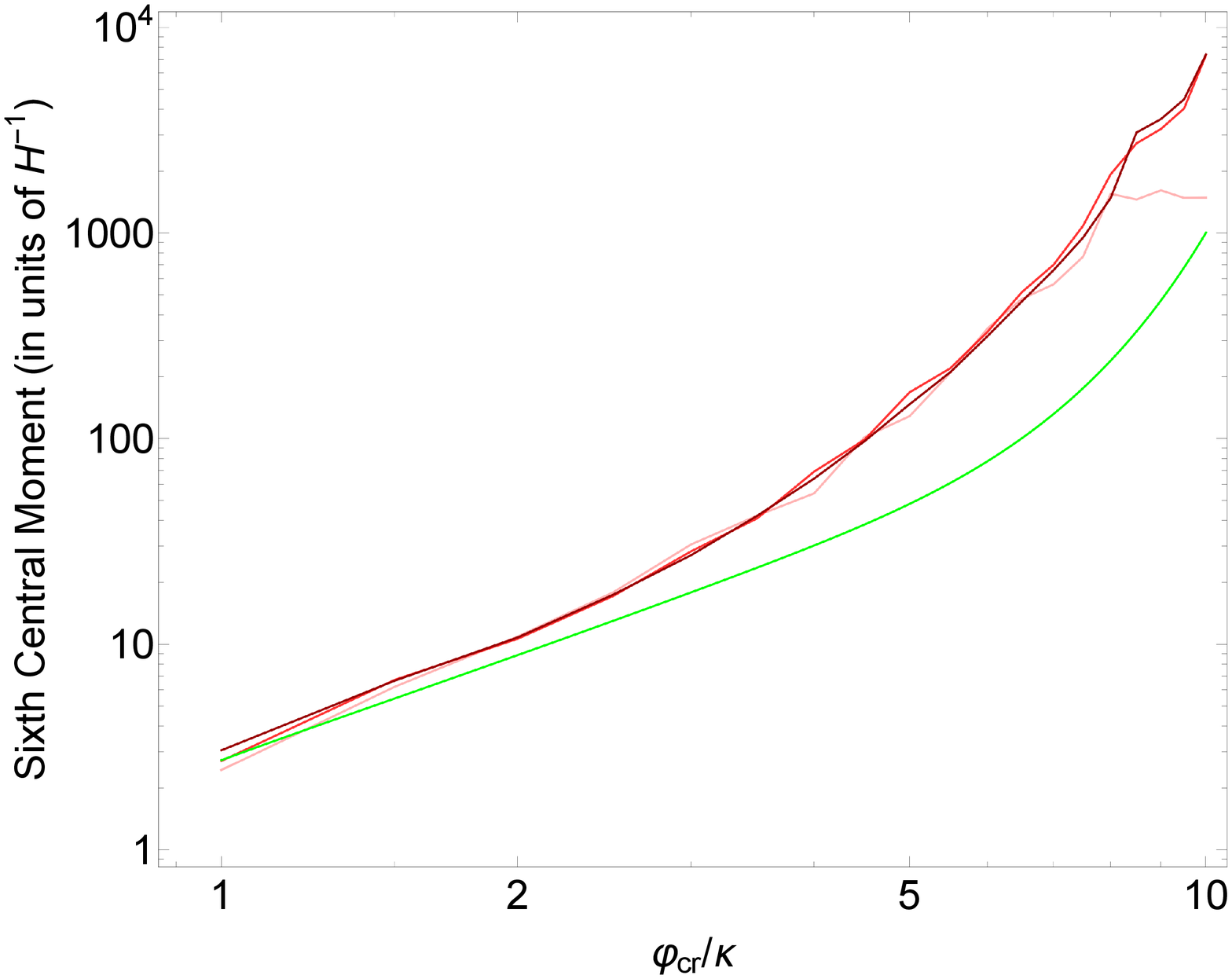}
\caption{Central moments of inflating regions for network-I for $12$, $15$, and $18$ 2-foldings (in increasing darkness) with the rough prediction from combinatorics (and kernel propagation) for comparison (green curve). Top panel is second central moment, middle panel is fourth central moment, and bottom panel is sixth central moment.}
\label{Moments1}
\end{figure}

\begin{figure}[h]
\centering
\includegraphics[width=0.95\columnwidth]{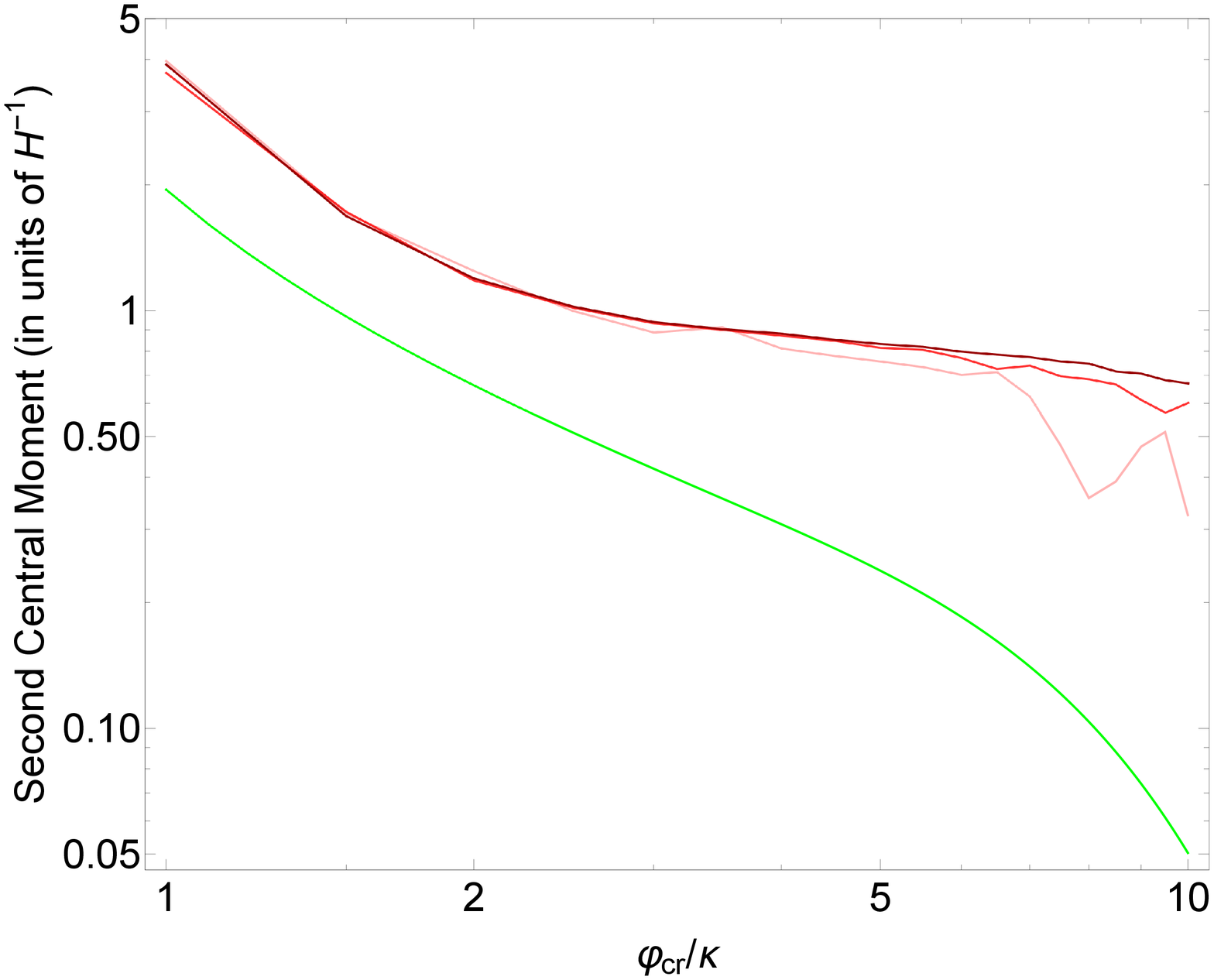}\vspace{\sz}
\includegraphics[width=0.95\columnwidth]{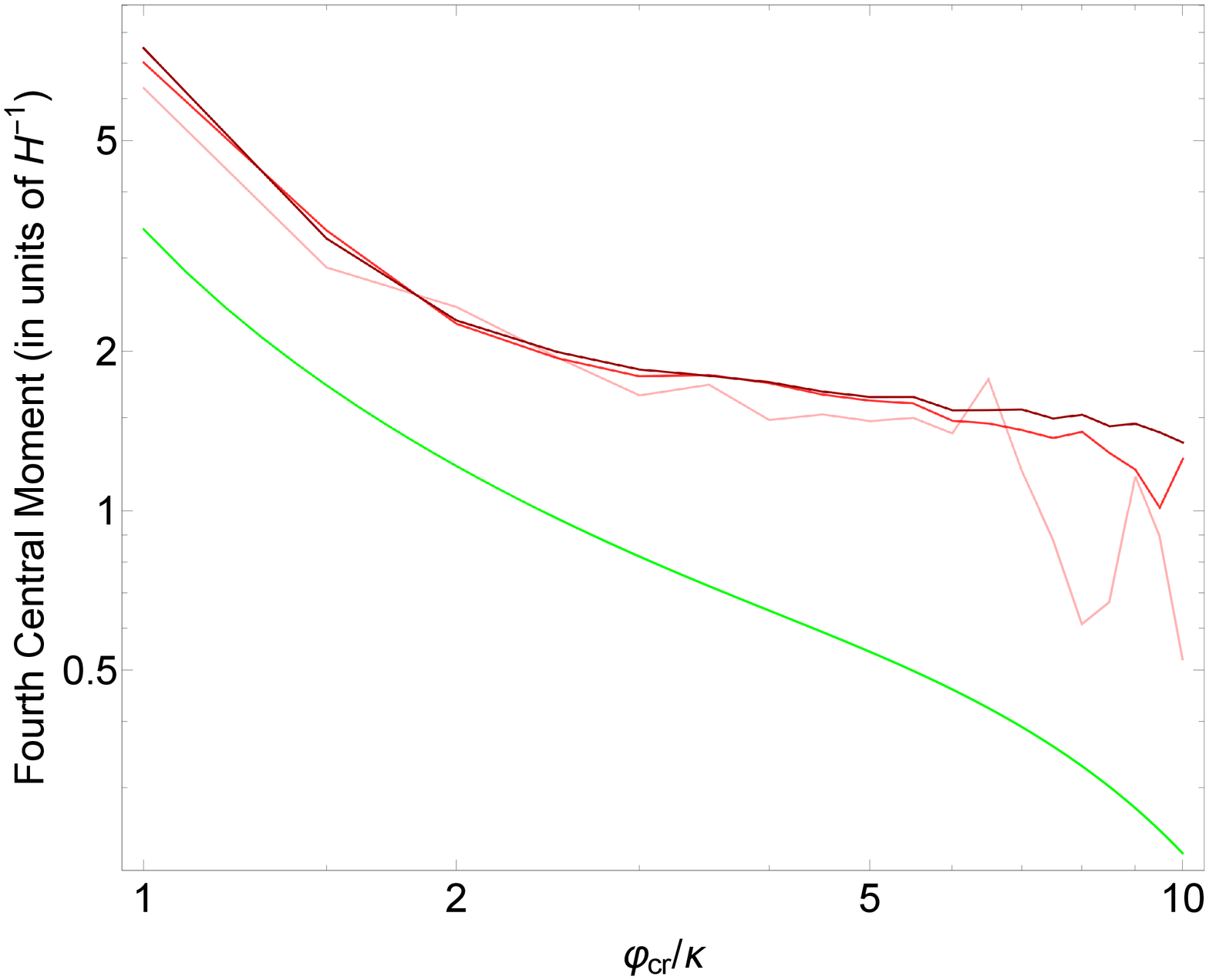}\vspace{\sz}
\includegraphics[width=0.95\columnwidth]{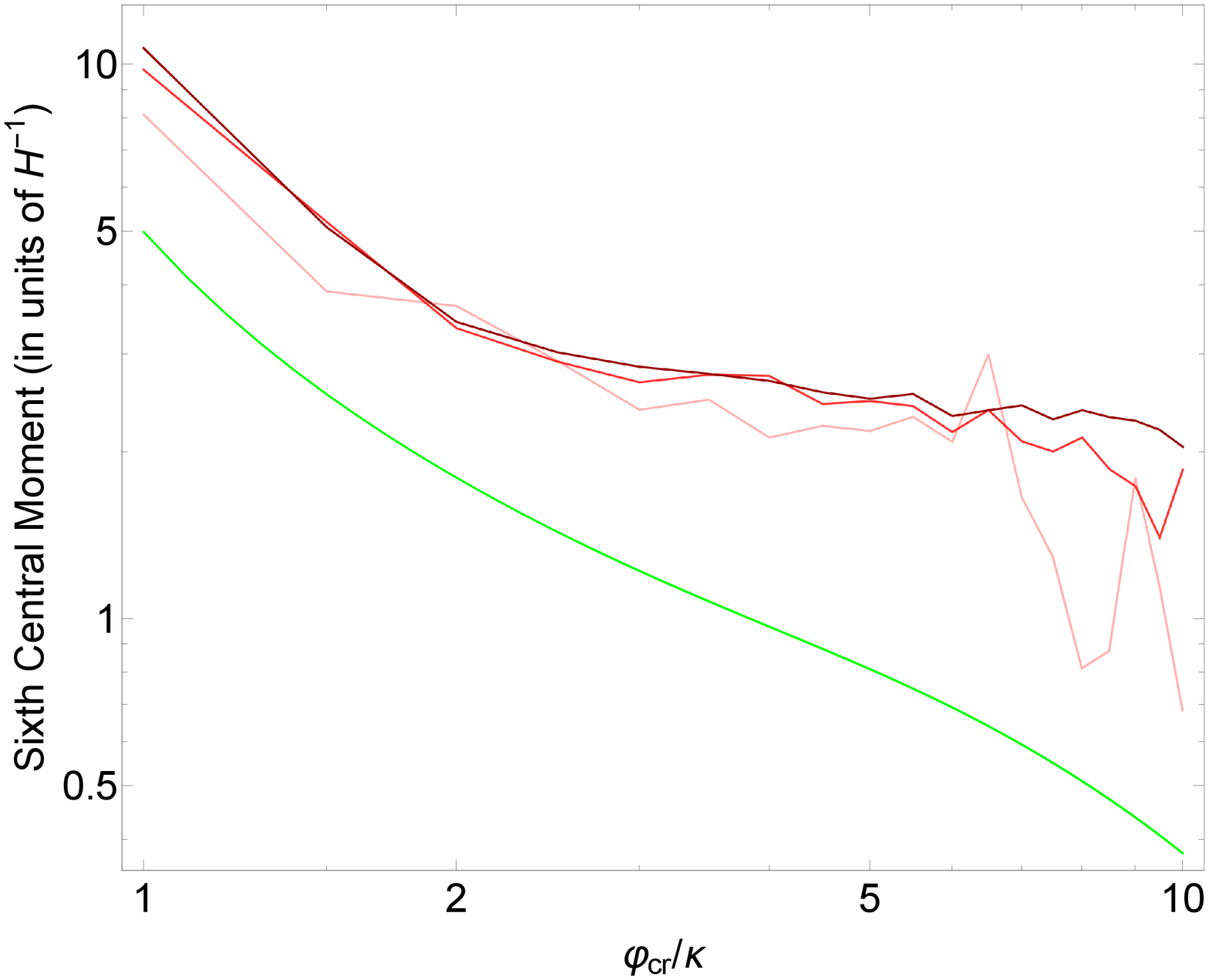}
\caption{Central moments of crunched regions for network-I for $12$, $15$, and $18$ 2-foldings (in increasing darkness) with the rough prediction from combinatorics (and kernel propagation) for comparison (green curve). Top panel is second central moment, middle panel is fourth central moment, and bottom panel is sixth central moment.}
\label{Moments1crunched}
\end{figure}

Figs.~\ref{Moments1} shows some central moments for network-I for $12, 15$, and $18$ 2-foldings (in increasing darkness) for lengths of inflating regions. Since the moments don't grow with more 2-foldings, it is suggestive that the tail of the distribution falls off faster than a power law. Results for network-II are similar. Similarly, Figs. ~\ref{Moments1crunched} shows some central moments for network-I for $12, 15$, and $18$ 2-foldings (in increasing darkness) for lengths of non-inflating regions. Note that the apparent increment in moments as we go higher in 2-foldings is suggestively due to not having enough statistics in the tail of the distribution for smaller 2-foldings.


\end{document}